\documentclass[epj,nopacs]{svjour}
\usepackage[switch,displaymath]{lineno} 
\usepackage{latexsym}  \usepackage{graphicx}  \usepackage{esvect}
\usepackage{amsmath}  \usepackage{mathabx} 
\usepackage{hepparticles} % particle names
\usepackage{hepnames} % shortcuts for lots of particle names
\usepackage[alsoload=hep]{siunitx}
\usepackage{mathtools}

\begin{document}
\title{High precision measurement of the associated strangeness production in proton proton interactions}
\titlerunning {High statistics measurement of the reaction pp $\rightarrow$ pK$^{+}\mathrm{\Lambda}$}
\subtitle{The COSY-TOF Collaboration}
\author{S.~Jowzaee\inst{1,7} \and
  E.~Borodina\inst{1}\and
  H.~Clement\inst{4,6} \and
  E.~Doroshkevich\inst{5,6}\thanks{current address: Institute for Nuclear Research Moscow 117312, Russia} \and
  R.~Dzhygadlo\inst{1}\thanks {current address: Hadron Physics I, GSI Helmholtzzentrum f\"{u}r Schwerionenforschung GmbH}\and
  K.~Ehrhardt\inst{5,6} \and
  W.~Eyrich\inst{4} \and
  W.~Gast\inst{1} \and
  A.~Gillitzer\inst{1} \and
  D.~Grzonka\inst{1} \and
  F.~Hauenstein\inst{1,4} \and
  P.~Klaja\inst{1,4} \and
  L.~Kober\inst{4} \and
  K.~Kilian\inst{1} \and
  M.~Krapp\inst{4} \and
  M.~Mertens\inst{1}\thanks {current address: Universit\"{a}t Duisburg-Essen 45141 Essen, Germany}\and
  P.~Moskal\inst{7}\and
  J.~Ritman\inst{1,2,8} \and
  E.~Roderburg\inst{1}\thanks{corresponding author e.roderburg@fz-juelich.de} \and
  M.~R\"{o}der\inst{1}\thanks {current address: Corporate Development, Forschungszentrum J\"{u}lich, 52428 J\"{u}lich, Germany} \and
  W.~Schroeder\inst{9} \and
  T.~Sefzick\inst{1} \and
  J.~Smyrski\inst{7} \and
  P.~Wintz\inst{1} \and
  P.~W\"{u}stner\inst{3}}

\institute{
  Institut f\"{u}r Kernphysik, Forschungszentrum J\"{u}lich, 52428 J\"{u}lich, Germany \and %1
  J\"{u}lich Aachen Research Allianz, Forces and Matter Experiments  (JARA-FAME) \and %2
  Zentralinstitut f\"{u}r Engineering, Elektronik und Analytik, 52428 J\"{u}lich, Germany \and %3
  Friedrich-Alexander-Universit\"{a}t Erlangen-N\"{u}rnberg, 91058 Erlangen, Germany \and %4
  Physikalisches Institut der Universit\"{a}t T\"{u}bingen, Auf der Morgenstelle 14, 72076 T\"{u}bingen, Germany  \and %5
  Kepler Center for Astro and Particle Physics, University of T\"{u}bingen, Auf der Morgenstelle 14, 72076 T\"{u}bingen, Germany \and %6 
  Institute of Physics, Jagellonian University, PL-30-348 Cracow, Poland \and %7
  Experimentalphysik I, Ruhr-Universit\"{a}t Bochum, 44780 Bochum, Germany \and %8
  Corporate Development, Forschungszentrum J\"{u}lich, 52428 J\"{u}lich, Germany %9
}
\date{\today}
\authorrunning{The COSY-TOF Collaboration}

\abstract
{ A new  high precision measurement of the reaction pp $\rightarrow$
  pK$^{+}\mathrm{\Lambda}$ at a beam momentum of
  $2.95\,\mathrm{GeV}/\mathrm{c}$ with more than 200\,000 analyzed
  events allows a detailed analysis of differential observables and
  their inter-dependencies. Correlations of the angular distributions
  with momenta are examined.  The invariant mass distributions are
  compared for different regions in the Dalitz plots.  The cusp
  structure at the $\mathrm{N\Sigma}$ threshold is described with the
  Flatt\'{e} formalism and its variation in the Dalitz plot is
  analyzed.}

\PACS{
      {13.75.-n} {Hadron-induced low- and intermediate-energy reactions and scattering (energy $\leq$ 10~GeV)} \and
      {13.75.Ev} {Hyperon-nucleon interactions}\and
      {25.40.Ve} {Other reactions above meson production thresholds (energies $>$ 400 MeV)}
     }
 
\maketitle
%\linenumbers
%\modulolinenumbers[2]

%======================================================================================================================================
\section{Introduction}

%...................................................................................
The associated strangeness production of kaon and lambda is the
energetic lowest possibility of creating particles with open
strange\-ness in nucleon nucleon collisions. This elementary reaction
is of fundamental interest as it involves the dissociation of an $\mathrm{s}\bar{\mathrm{s}}$
quark pair into two hadrons with $\mathrm{s}=1$ and
$\mathrm{s}=-1$. 

The creation or dissociation of the $\mathrm{s}\bar{\mathrm{s}}$ quark pair is described in
different models by the exchange of a meson between the incoming
nucleons. In most models kaon or pion exchange is assumed, but this
approach can be extended to  $\mathrm{\eta}$, $\mathrm{\rho}$, and $\mathrm{\omega}$ mesons and to the
strange K* mesons  \cite {Ferrari1968a,Laget1991,Sibirtsev2006,Shyam2006,Liu2006}.  To separate the contributions of the
strange and non strange meson exchange experimentally, two
possibilities are proposed: First significant differences of the
analyzing power and the spin transfer coefficient for both exchanges
are predicted \cite{Laget1991}. Secondly the evidence of nucleon resonances in
the lambda kaon subsystem can only be explained by non strange meson
exchange \cite{Sibirtsev1998b}.

The associated strange\-ness production in proton proton collisions has
been extensively studied by the DISTO collaboration \cite
{Balestra1999a}, the COSY-TOF collaboration \cite {AbdEl-Samad2010},
\cite{Abdel-Bary2010a}, close to threshold by the COSY11 collaboration
\cite {Rozek2006}, and at a higher beam momentum by the HADES
collaboration \cite {Fabbietti2013}. In spite of this large
experimental data base no conclusive solution has been found on the
exchange mechanism. While measurements of the spin transfer
coefficient of the DISTO experiment indicate kaon exchange, COSY-TOF
measurements yield a large N* contribution \cite {AbdEl-Samad2010}, which is a sign for
dominant pion exchange.

Apart from the reaction mechanism itself, the associated strange\-ness
production offers powerful tools to study a) the decay of N* resonances
into channels with strange\-ness, b) the hyperon nucleon interaction, and
c) the coupled channel effect of (N$\mathrm{\Lambda}$~$\leftrightarrow$~N$\mathrm{\Sigma}$)
at the N$\mathrm{\Sigma}$ threshold.

a) The first evidence for the creation of N* resonances in the associated
strange\-ness production were found by comparing the Dalitz plot
distributions measured with the COSY-TOF experiment \cite {AbdEl-Samad2010,Abdel-Samad2006a}
with predictions, which assume a strong impact of the resonances
N(1650), N(1710), and N(1720). Further insight to the
weight of these resonances and their branching ratio to the K$\mathrm{\Lambda}$ channel
can be obtained by a partial wave analysis, with precise data --
including polarization observables -- as input.  Therefore, a combined
analysis of all existing data with the Bonn-Gatchina partial wave
analysis program \cite {Anisovich2007} has been star\-ted \cite
{Muenzer2015}. First results of the partial wave analysis of HADES data are given in  \cite {Agakishiev2015}.

b) The invariant mass distributions of the associated strange\-ness
production allows the interaction of the involved particles to be studied as a function
 of the relative energy in the range from zero to several hundred
MeV. Especially the measurement of the nucleon hyperon invariant mass
distributions starting from zero interaction energy enables the
determination of the nucleon hyperon scattering length \cite {Gasparyan2005,Xie2011b}.
The effective p$\mathrm{\Lambda}$ scattering length has been determined
from COSY-TOF data \cite{Roeder2013}. A precise knowledge of the
N* resonances is needed, as they influence the p$\mathrm{\Lambda}$ invariant mass
spectrum by reflections of the K$\mathrm{\Lambda}$ channel, thus limiting the
precision of the scattering length determination.

c) The last tool provided by the associated strange\-ness production is to
study the coupled channel effect at the threshold of N$\mathrm{\Sigma}$ in the
nucleon hyperon invariant mass distribution. The cusp structure at the
threshold was discovered in 1961 by a bubble chamber measurement of K$^{-}$
absorption in deuteron leading to p$\mathrm{\Lambda}\mathrm{\pi^{-}}$ \cite {Dahl1961b}.  
This was
confirmed by higher statistic measurements of the same reaction \cite {Tan1969b,Braun1977}.
The measurements were described by different models and
calculations \cite {Toker1981,Badalyan1982,Torres1986}. As for the pK$\mathrm{\Lambda}$ final
state both thresholds of N$\mathrm{\Sigma}$ and K$\mathrm{\Sigma}$ have to be exceeded, the
cusp effect can be observed for beam momenta above $2.7\,\mathrm{GeV}/\mathrm{c}$.  First
evidences for a cusp in this final state at beam momenta of $2.75\,\mathrm{GeV}/\mathrm{c}$
 and $2.85\,\mathrm{GeV}/\mathrm{c}$ were reported of measurements with the COSY-TOF
experiment \cite {Abdel-Samad2006a}. A detailed examination of the
cusp was performed with data measured at a beam momentum of $3.05\,\mathrm{GeV}/\mathrm{c}$
\cite{AbdEl-Samad2013}.

In addition to the fundamental processes which can be studied with
the associated strange\-ness production, this elementary reaction is the
basis of more complex reactions, which are under investigation and
which need the precise knowledge of its observables. Among these
complex reactions are the properties of hypernuclei \cite {Gibson1995}, the possible
existence of ppK cluster or bound states \cite{Agnello2005c}, and the prediction of
strange\-ness enhancement in heavy ion interactions as a sign for the
quark gluon plasma \cite{Rafelski1982}. Even information of the associated strangeness 
production measured at other kinematic regions can contribute to a better understanding, as
they can help to describe the involved nucleon resonances and threshold effects, which have to 
be subtracted in order to identify these higher order effects. 

The purpose of this paper is to present the analysis of a high
precision measurement of the reaction pp $\rightarrow$ pK$^{+}\mathrm{\Lambda}$      in order to
increase the existing data base in this momentum region by a factor of
5. Special emphasis is placed on 
the description of the threshold effect.
Angular distributions, Dalitz plots, and invariant mass
distributions are given and examined. The polarization variables will be presented
in a forthcoming paper.

%======================================================================================================================================
\section{Experimental setup}
\begin{figure}[thb]
	\begin {center}
	\includegraphics[width=.5\textwidth]{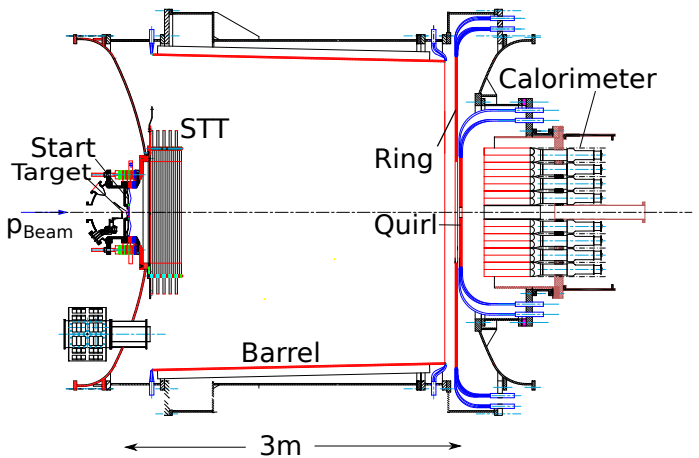}
	\caption{\label{experiment} Side view of the COSY-TOF 
          spectrometer. In beam direction the start counter (Start), 
          the straw tube tracker (STT), the barrel scintillators, the 
          inner ring (Quirl), the outer ring (Ring), and the 
          Calorimeter are shown.
          All detectors and the liquid hydrogen target are located inside the vacuum vessel.
        }  
        \end {center}
\end{figure}

The COSY-TOF experiment is a non-magnetic spectro\-meter, which is situated
at an external beam-line of the accelerator COSY in the research center
J\"{u}lich. It consists of 
a cylindrical vacuum vessel of $\SI{3.5}{\metre}$ length and $\SI{3}{\metre}$ dia\-meter (see fig.~\ref {experiment}).
The pressure inside the vessel is lower than $\SI{6e-4}{\hecto\pascal}$.  

\subsection{Scintillation trigger detectors}
At the inner walls of the cylinder part of the vessel 96 plastic
scintillator bars are attached, which have a thickness of
$\SI{15}{\milli\metre}$, a width of ca. $\SI{10}{\centi\metre}$ and a
length of $\SI{285}{\centi\metre}$.  They are arranged to form a
slightly conical barrel of about $\SI{300}{\centi\metre}$ in
dia\-meter. The barrel scintillator is described in ref. \cite
{Bohm2000}.  The downstream end-cap of the vessel is covered by three
layers of $\SI{5}{\milli\metre}$ thick plastic scintillators, which
are arranged in two rings, the inner one ranging from
$\SI{2.4}{\centi\metre}$ to $\SI{58}{\centi\metre}$ radius, the outer
one from $\SI{54}{\centi\metre}$ to $\SI{150}{\centi\metre}$. One of the layers is
composed of 48 (inner ring) and 96 (outer ring) wedge-shaped
scintillators. Each of the other two layers consists of 24 (inner
ring) and 48 (outer ring) elements, formed as Archimedian spirals. The
spirals change the direction of rotation from one to the other layer,
therefore, 1104 (inner ring) and 2304 (outer ring) pixels can be
defined by coincidences. The inner ring is described in ref. \cite
{Dahmen1994}. The scintillators are read out by XP2020
photo-multipliers.  Downstream, the inner ring counter is followed by
a scintillator calorimeter.  The data of the calorimeter are not used
for this measurement.

\subsection{Target and beam defining detectors}
The liquid hydrogen target is located in the entrance part of the
vacuum vessel. It is a cylindrical cell with a dia\-meter of
$\SI{6}{\milli\metre}$ and a length of $\SI{4}{\milli\metre}$, which
consists of $\approx\,\SI{4}{\micro\metre}$ thick copper walls in the
barrel part and of $\SI{0.9}{\micro\metre}$ thick Mylar window foils in
the entrance and exit.  In order to reduce rest gas deposition onto
the target windows, the pressure surrounding the target cell is kept
below $\SI{2e-7}{\hecto\pascal}$.  A scintillation start counter is
located at a distance of about $\SI{15}{\milli\metre}$ downstream to
the target center. It is made from two $\SI{1}{\milli\metre}$ thick
discs, with an outer radius of $\SI{75}{\milli\metre}$ and a central
beam hole of $\SI{1.5}{\milli\metre}$ radius. Each disk is divided
into 12 wedge-shaped parts, which are individually read out by
Hamamatsu R1450 photo-multipliers.  A double-sided silicon detector
disk with a thickness of $\SI{300}{\micro\metre}$ and a central beam
hole of $\SI{2.75}{\milli\metre}$ radius is located
$\SI{26}{\milli\metre}$ downstream. The data of this detector are only
applied for pK$\mathrm{\Sigma}^+$ analysis, it will be described in a
forthcoming paper on the $\mathrm{\Sigma}^+$ results.  In order to
veto beam particles which hit directly the start counter, a
scintillation counter with a hole of $\SI{1}{\milli\metre}$ radius is
installed $\SI{50}{\milli\metre}$ upstream of the target. Further
upstream at $\SI{50}{\centi\metre}$ and $\SI{100}{\centi\metre}$ two
veto counters with central holes of 6 and $\SI{15} {\milli\metre}$ diameter 
detect beam halo particles which are around the beam core. The total
beam rate and its x- and y-intensity distribution is measured with a
fiber hodoscope located $\SI{1}{\metre}$ downstream of the
spectrometer exit. The target, start and veto counters are aligned to
the center of the beam line with a precision in the order of
$\SI{0.1}{\milli\metre}$ with an optical telescope, which can be
placed $\SI{2}{\metre}$ upstream of the target into the beam line. The
beam is focused onto the target with a profile which has a full width
at half maximum of about $\SI{1}{\milli\metre}$. Due to precise tuning
of the beam line magnets the maximum rate in the veto counters is less
than 1\% of the total beam rate.
\subsection{Tracking detector}
The main detector system is the central straw tube tracker (STT).  The
STT is placed about $\SI{24}{\centi\metre}$ behind the target inside
the vacuum tank. It consists of 2704 straw tubes, which are arranged
in 13 double layers and fixed in six orientations with an angle of
\ang{60} to each other in order to enable 3D track reconstruction. A
single straw has a length of $\SI{1050}{\milli\metre}$,
$\SI{10}{\milli\metre}$ dia\-meter and $\SI{30}{\micro\metre}$ wall
thickness. The active detector volume of the 13 double layers is about
$\SI{1}{\metre}$ in dia\-meter and $\SI{24}{\centi\metre}$ in
depth. An inner beam hole in each double layer with
$\SI{15}{\milli\metre}$ dia\-meter avoids beam interactions with the
detector. Each straw tube is made of aluminized Mylar film and filled
with Ar + CO$_2$ (80~:~20) gas mixture at a pressure of
$\SI{1.2}{\bar}$. The straws are self-supporting by the over pressure
of the chamber gas. Therefore, sufficient mechanical stability is
provided, despite the low material budget of $X/X_0 \approx 1\%$ for
26 layers of straws. The anode is a $\SI{20}{\micro\metre}$ thick gold
plated tungsten-rhenium wire which is stretched along the straw axis
and held at a potential of $+\SI{1820}{\volt}$ \cite{Wintz2004}.

The arrangement of the target and detectors facilitates a volume 
in the z-direction from the target to the beginning of the straw detector, which is 
-- apart from the start counter and silicon counter -- free of
any material. Therefore, a secondary vertex in this volume indicates
with highest probability a $\mathrm{\Lambda}$ or a K$_{\mathrm S}^0$ decay. 

\subsection{Trigger}
The trigger is defined by signals of the start counter, the barrel scintillators 
and the end-cap scintillators. At least one signal of the start counter
and at least 4 hits in the barrel and end-cap are required. A
hit in the ring counters is defined by coincident signals in two of
the three layers and a hit in the barrel counter is defined by a
coincidence signal of the photo-multipliers, which are situated on both ends of a barrel
scintillator bar.  The coincidence width is $\SI{100}{\nano\second}$, which takes into
account the different time of flight of particles between the start
and stop counters and of different light paths in the scintillators
and light-guides.  In order to facilitate the rate of $\approx$ 2000
events/s written to disk, only the STT data were read out, as this
read out electronics has a dead time of $\SI{10}{\micro\second}$. The dead time of
the ADC and TDC readout for the scintillators is in the order of
$\SI{200}{\micro\second}$. Therefore, data which include the ADC and TDC information
of the scintillators are recorded for controlling the trigger system
for only about 5\% of the beam time.

%======================================================================================================================================

\section{Analysis}

\subsection{Event Reconstruction}
The analysis of the pK$\mathrm{\Lambda}$ events is based on the tracking
information obtained from the straw tube tracker. The determination of
the correlation between drift time and the track distance to the anode
is performed in two steps \cite{Jowzaee2013,Jowzaee2014}: In the first step the correlation
is obtained by an integration over all
measured drift times and by assuming a homogeneous illumination of the
straw tube by particle tracks. In the second step the position in the straw tube
is calculated from reconstructed tracks, which are crossing this tube. 
With this information 
the correlation between drift time and distance
to the center is recalculated. The spatial resolution of the STT is
$\sigma = \SI[parse-numbers=false]{(140 \pm 10)}{\micro\metre}$ at $\SI{2.5}{\milli\metre}$ distance to the anode \cite {Jowzaee2014}.

The main part of the program which extracts the tracks from the straw
tube isochrones is contained in the tof\-Straw library
\cite{Castelijns2006} and described in detail in \cite {Jowzaee2014},
\cite{Hauenstein2014}, and \cite{Roeder2011}.  The primary and decay
vertices are submitted to a geometrical fit, which includes the
information, that the $\mathrm{\Lambda}$ decay plane has to contain the primary
vertex.  The events, for which the fit converges, are submitted to a
kinematic fit by taking the geometrical fit results as start
values. The kinematic fit applies momentum and energy conservation,
the masses of the $\mathrm{\Lambda}$ and its decay particles are directly used
as input, as the $\mathrm{\Lambda}$ is identified by the secondary vertex and
the decay proton has always the smaller angle to the $\mathrm{\Lambda}$
direction.  The assignment of the proton and kaon masses to the
primary tracks is done by applying the kinematic fit for each of the
two solutions, the result with the lower $\chi^2$ is chosen. Events
with a $\chi^2$/NDF of less than 5, a decay length in the rest system
of the $\mathrm{\Lambda}$ of larger than $\SI{2}{\centi\metre}$ and an angle of the decay proton
to the $\mathrm{\Lambda}$ direction of larger than \ang{2.5} are evaluated as
pK$\mathrm{\Lambda}$ events.

\subsection{Monte Carlo calculations}
The corrections for the detector acceptance and the reconstruction
efficiency are done with a Monte Carlo program (GEANT 3.2
\cite{Brun1984}) with phase-space distributed events as input.
The corrections were controlled with a Monte Carlo program with an input, which simulates
the measured proton cm angular distribution. No significant changes of the resulting
angular distributions were found.
The combined acceptance and reconstruction efficiency of pp $\rightarrow$
pK($\mathrm{\Lambda} \rightarrow$ p$\mathrm{\pi}^{-}$) amounts to
$\SI[parse-numbers=false]{(22.5 \pm 1)}{\percent}$.  The primary
vertex resolution is $\sigma(\mathrm{x},\mathrm{y}) =
\SI[parse-numbers=false]{(0.5 \pm 0.1)}{\milli\metre}$ and
$\sigma(\mathrm{z}) = \SI[parse-numbers=false]{(1.5 \pm
  0.2)}{\milli\metre}$. The transverse resolution of the secondary
vertex is in the same range as of the primary one, while the
longitudinal resolution is worse: $\sigma(\mathrm{z}) =
\SI[parse-numbers=false]{(2.0 \pm 0.5)}{\milli\metre}$.  The
resolution of the p$\mathrm{\Lambda}$ invariant mass is
$\sigma(m_{\mathrm{p\Lambda}}) = (1.1 \pm
0.3)\,\mathrm{MeV}/\mathrm{c}^{2}$, the resolution of the K$\mathrm{\Lambda}$ 
and pK invariant masses are 
$\sigma(m_{\mathrm{K\Lambda}}) = (1.1 \pm 0.3)\,\mathrm{MeV}/\mathrm{c}^{2}$ and
$\sigma(m_{\mathrm{pK}}) = (1.0 \pm 0.3)\,\mathrm{MeV}/\mathrm{c}^{2}$
 \cite  {Jowzaee2014,Roeder2011}.

\subsection{Background}
\begin{figure}[htpb]
	\begin {center}
	\includegraphics[width=.5\textwidth]{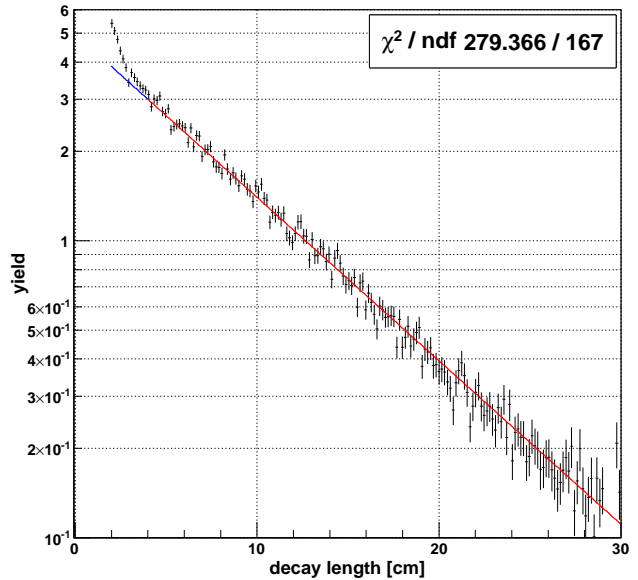}
	\caption{\label{decaylength} $\mathrm{\Lambda}$ decay length distribution in the
          $\mathrm{\Lambda}$ rest system.  The data are corrected for
          acceptance and efficiency. The fit  with $p_{0}\cdot \exp(-x/7.89\,\mathrm{cm})$ is applied in the range
          between $\SI{4}{\centi\metre}$ and $\SI{30}{\centi\metre}$ (red line). The prolongation 
          between $\SI{2}{\centi\metre}$ and $\SI{4}{\centi\metre}$ is shown
          with the blue line.
        }  
        \end {center}
\end{figure}

One source of background are multi pion events.  They can produce
spurious secondary vertices if tracks are reconstructed erroneously.
These spurious secondary vertices are expected to be close to the
target, as the density of tracks is highest in this region.  These
background events can be identified as deviation of the measured from
the calculated $\mathrm{\Lambda}$ decay length (fig.~\ref
{decaylength}). The normalized data are fitted with $f(x) = p_{0}\cdot
\exp(-x/\mathrm{c}\tau_{0})$ in the range between
$\SI{4}{\centi\metre}$ and $\SI{30}{\centi\metre}$.  $p_{0}$ is the
fit coefficient and c$\tau_{0}$ is the literature value of the
$\mathrm{\Lambda}$ decay length (7.89~cm) \cite{Olive2014}.  The
background events are concentrated in the first centimeters of the
decay length. The excess of the data compared to the prolongation of
the fit curve between $\SI{2}{\centi\metre}$ and
$\SI{4}{\centi\metre}$ adds up to 5\% of the total events.

The $\mathrm{\Sigma}^{0}$ background events caused by
$\mathrm{\Sigma}^{0} \rightarrow \gamma \mathrm{\Lambda}$ cannot be
detected by geometry information, as the event topology is similar to
the pK$\mathrm{\Lambda}$ events. The suppression of these events by
the $\chi^{2}$ cut of the kinematic fit is studied with a Monte Carlo
sample of pp $\rightarrow$ pK$\mathrm{\Sigma}^{0} (\rightarrow
\mathrm{\Lambda} \gamma)$.  The reconstruction efficiency of this
sample is $\SI{8.1}{\percent}$, which has to be compared to the
reconstruction efficiency of pp $\rightarrow$ pK$\mathrm{\Lambda}$
which amounts $\SI{22.5}{\percent}$.  Including the ratio of the cross
sections of $\SI{3.1}{\micro\barn} / \SI{21.8}{\micro\barn}$ \cite
{Abdel-Bary2010a} a contamination of $\SI{5}{\percent}$ of
$\mathrm{\Sigma}^{0}$ events is obtained. The analysis of the Monte
Carlo data of pK$\mathrm{\Sigma}^{0}$ shows, that these events are
shifted by the kinematic fit to the backward region of the
$\mathrm{\Lambda}$ and kaon cm scattering angles. For
pK$\mathrm{\Lambda}$ events this effect is not observed.
%======================================================================================================================================
\section{Results}
\subsection{Angular distributions in the cm system}
\begin{figure}[tb]
	\begin {center}
	\includegraphics[width=.5\textwidth]{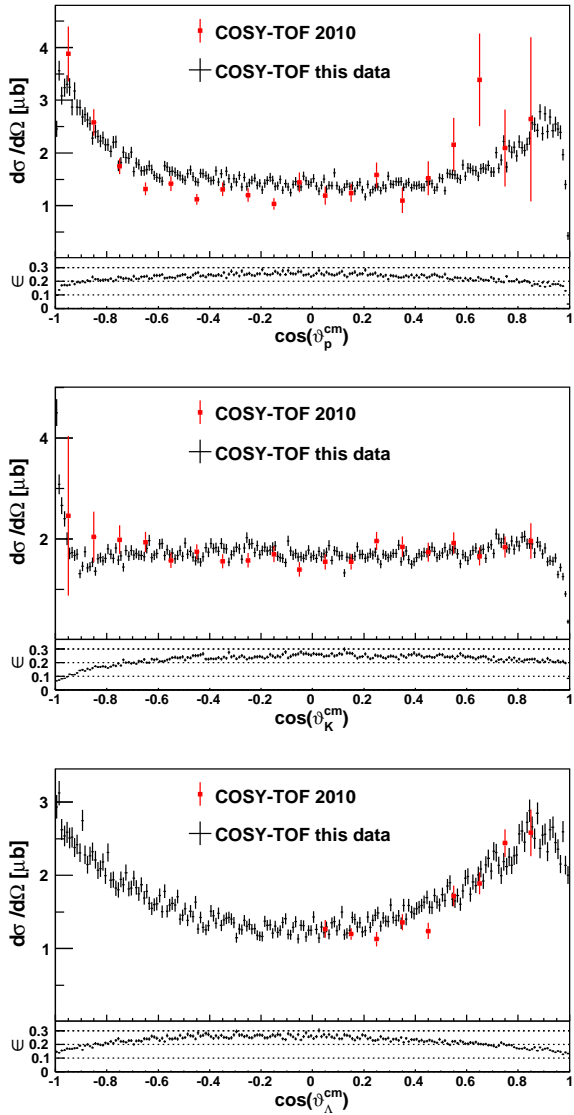}
	\caption{\label{angular-distribution-pcm} The angular
          distributions in the cm system are shown for the proton
          (upper figure), kaon (middle), and $\mathrm{\Lambda}$ (bottom).  The
          results from a former COSY-TOF measurement \cite {Abdel-Bary2010a} 
          are given as red squares. Beneath each
          distribution the detector acceptance and reconstruction
          efficiency $\mathrm{\epsilon}$ is shown.
        }  
        \end {center}
\end{figure}

The angular distributions in the center of mass system are shown in
fig.~\ref{angular-distribution-pcm}. 
 The data are corrected for the detector acceptance and reconstruction
 efficiency and are normalized to the total cross section of
 $\SI{21.8}{\micro\barn}$, which is determined in a former COSY-TOF
 measurement \cite {Abdel-Bary2010a}. As target and projectile are
 identical particles, which cannot be distinguished in the~cm system,
 these distributions have to be symmetric to
 $\cos\vartheta^{\mathrm{cm}} = 0$.  Small deviations from symmetry
 are existing for all three distributions: At backward angles there is
 a surplus of events, which is mainly due to the
 pK($\mathrm{\Sigma}^{0}\rightarrow \mathrm{\Lambda} \gamma$)
 background. At extreme forward angles ($\cos\vartheta^{\mathrm{cm}}
 \ge 0.95$) of the proton and kaon distributions a deficit of events
 is found.  The affected range corresponds to scattering angles in the
 lab system of smaller than \ang{5} - \ang{6} degrees. It is supposed
 that the deficit of events is caused by an efficiency drop in the
 cones of the start counter, which cannot be exactly described by the
 Monte Carlo program.

While the angular distribution of the heavier reaction products, p and
$\mathrm{\Lambda}$, exhibit distinct deviations from uniformity, the
distribution of the lighter kaon is nearly constant. This behavior is
examined in detail in section \ref{detailedAngularDistribution}.

\subsection{Gottfried-Jackson angular distributions}

\begin{figure}[tb]
	\begin {center}
	\includegraphics[width=.5\textwidth]{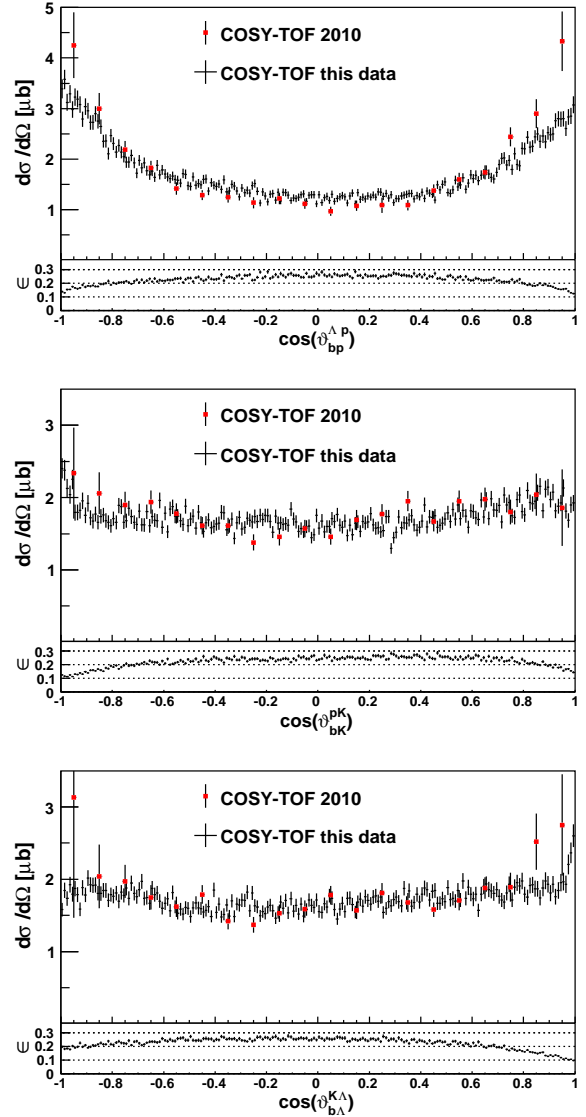}
	\caption{\label{gfj-angular-distribution} 
          The angular distribution in the  p$\mathrm{\Lambda}$ Gottfried-Jackson frame is 
          shown in the upper figure. Middle:  pK Gottfried-Jackson frame, 
          bottom: K$\mathrm{\Lambda}$ Gottfried-Jackson frame.         
          The results from a former COSY-TOF measurement \cite
          {Abdel-Bary2010a} are given as red squares. Beneath each
          distribution the detector acceptance and reconstruction
          efficiency $\mathrm{\epsilon}$ is shown.
        }  
        \end {center}
\end{figure}

The angular distributions in the three two particle subsystems of the
 final state pK$\mathrm{\Lambda}$ are constructed by boosting the
momenta into the rest system of the two particle system. The angle
between the boosted beam vector and one of the particles of the two
particle system is the Gottfried-Jackson (GJ) angle. It is denoted with
$\vartheta^{\mathrm{ij}}_{\mathrm{bi}}$, where i, j are the particles of the two particle
rest system and b stands for the boosted beam direction.  The GJ angular
distributions are shown for the rest systems of $\mathrm{\Lambda}$p, pK, and
K$\mathrm{\Lambda}$ in fig.~\ref{gfj-angular-distribution}.
In contrast to the cm distributions the GJ angular distributions do
not have to be symmetric to $\cos\vartheta^{\mathrm{ij}}_{\mathrm{bi}} =0$.

While only in the K$\mathrm{\Lambda}$ rest system higher angular momenta due to nucleon
resonances are expected, the GJ angular distributions exhibit the
highest anisotropy for the $\mathrm{\Lambda}$p rest system.  The GJ angular
distributions are examined in dependence on bins of the invariant
masses in section \ref{detailedGJ}.

\subsection{Dalitz plots}
\begin{figure}[htbp]
	\begin {center}
	\includegraphics[width=.5\textwidth]{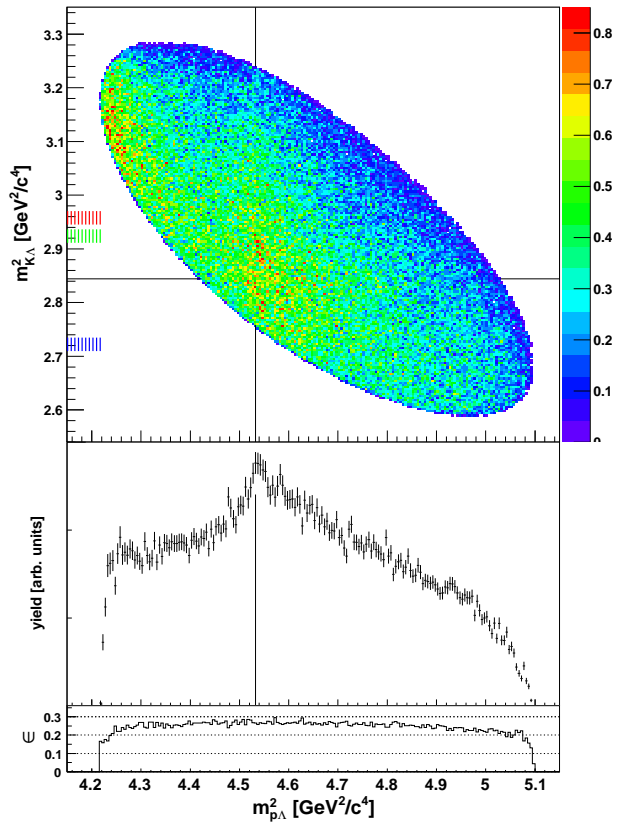}
	\caption{\label{dalitz-pl-kl} Top: The Dalitz plot of the
          measured data. In order to enhance the visibility of the
          structures the color code is truncated to 0.85 (less
          %dalitzplot 200 x 200 = 40 000 bins, belegt 1/3 : 12 * 100 * 3/ 40 000 = 0.1%
         than 0.1\% of the bins have an occupancy of larger than 0.85). The
          black horizontal line indicates the K$\mathrm{\Sigma}$ threshold, the
          vertical one the N$\mathrm{\Sigma}$ threshold.  The red area at the
          ordinate indicates the mass of the N$^*$(1720) resonance,
          the green area indicates the N$^*$(1710) resonance, and the
          blue area the N$^*$(1650) resonance.
          Middle: The projection on the squared invariant mass
          $\mathrm{p\Lambda}$ is shown. This projection is corrected by the
          combined detector and reconstruction efficiency, which is
          shown in the bottom frame. The N$\mathrm{\Sigma}$ threshold is
          indicated by the black line.   }	
        \end {center}

\end{figure}

The Dalitz plots are normalized by scaling the maximum channel content to 1.
The Dalitz plot of $m_{\mathrm{p\Lambda}}^{2}$ $versus$ $m_{\mathrm{K\Lambda}}^{2}$ is shown in
fig.~\ref {dalitz-pl-kl}. Two dominant structures, which are stretched
mainly in the vertical direction, are observed.  The first one is
located at low p$\mathrm{\Lambda}$ invariant masses and arises from the
p$\mathrm{\Lambda}$ final state interaction \cite {Sibirtsev2006}. The second structure located at
the N$\mathrm{\Sigma}$ threshold has its maximum intensity around 
 $m_{\mathrm{K\Lambda}}^{2} \approx 2.85\,\mathrm{GeV}^{2}/\mathrm{c}^{4}$ and
extenuates with rising K$\mathrm{\Lambda}$ invariant mass. 

\begin{figure}[htbp]
	\begin {center}
	\includegraphics[width=.5\textwidth]{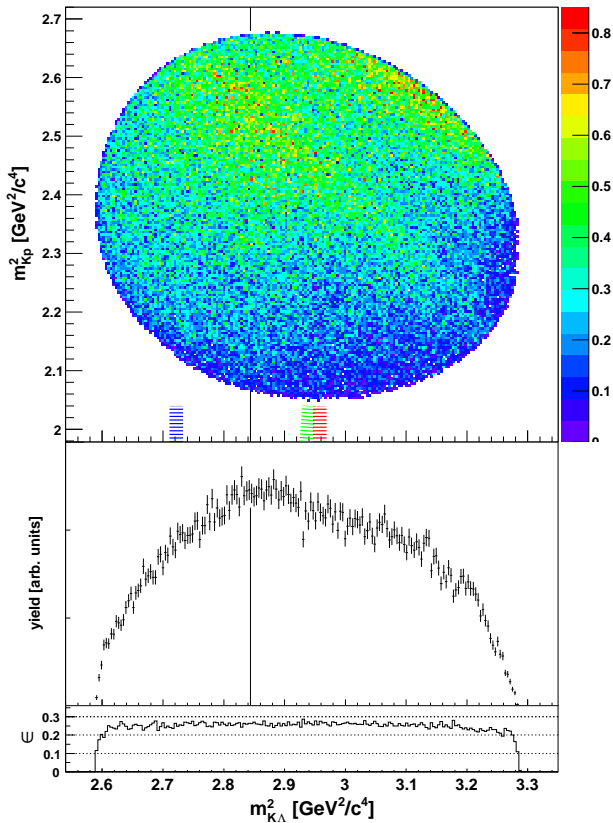}
	\caption{\label{dalitz-kl-pk} Top: The Dalitz plot of the
          measured data. In order to enhance the visibility of the
          %23 *100 *2 /40 000 = 0.1%
          structures the color code is truncated to 0.85 (less than 0.1\% of the bins
          have an occupancy of larger than 0.85). The black vertical line indicates the
          K$\mathrm{\Sigma}$ threshold.  The red area at the abscissa indicates the mass
          of the N$^*$(1720) resonance, the green area indicates the
          N$^*$(1710) resonance, and the blue area the N$^*$(1650)
          resonance. 
          Middle: The projection on the squared invariant mass K$\mathrm{\Lambda}$ is shown. 
          This projection is corrected by the combined detector and reconstruction efficiency,
          which is shown in the bottom frame. The K$\mathrm{\Sigma}$ threshold is indicated by the black line.}
	\end {center}

\end{figure}

The Dalitz plot of $m_{\mathrm{K\Lambda}}^{2}$ $versus$ $m_{\mathrm{Kp}}^{2}$ is shown in
fig.~\ref {dalitz-kl-pk}.  The two dominant structures of the previous
Dalitz plot are emerging as diagonal spread elevations. Again the
structure connected to the N$\mathrm{\Sigma}$ threshold exhibits a strong
intensity variation across the Dalitz plot.

No structures at the K$\mathrm{\Sigma}$ threshold or near the resonance masses
can be detected.  A detailed analysis of the invariant mass
distributions is given in section
\ref{detailedInvariantMassDistribution}.
%======================================================================================================================================
\section{Detailed examinations and discussions}
\subsection{CM angular distributions}\label{detailedAngularDistribution}
Former COSY-TOF measurements exhibit that the $l=2$ Legendre polynomial
coefficient, which is a measure for the contribution of higher partial waves, rises with
increasing beam momentum \cite {Abdel-Bary2010a}.  The data presented here
allow for the first time to extract the Legendre
coefficients not only from the integrated angular distribution but from
a series of angular distributions, each constructed for a bounded
range of the center of mass  momentum. 

For all reaction particles (p, K, $\mathrm{\Lambda}$) the magnitude of the
center of mass momentum of the particle is divided into 6 equidistant intervals. For each
momentum interval the angular distribution of the particle has been fit
according to the formula

\begin {equation} \label {legendreSum}  
  \frac{\mathrm{d}\sigma}{\mathrm{d}\cos\vartheta^{\mathrm{cm}}}=C\cdot(a_{0}P_{0}+a_{2}P_{2}(\cos\vartheta^{\mathrm{cm}})+a_{4}P_{4}(\cos\vartheta^{\mathrm{cm}}))
\end {equation}

$C$ is a normalization constant.  This formula describes the
differential cross section directly as a sum of Legendre polynomials $P_i$
by neglecting the interference terms.  Due to identical particles in
the initial state, the angular distributions in the center of mass
system have to be symmetric with respect to $\cos\vartheta^{\mathrm{cm}} = 0 $. Therefore,
the coefficients $a_{1}$ and $a_{3}$ are set to zero. With this
parametrization and the restriction to the maximum angular momentum of $l =2$, 
the coefficient $a_2$ contains products of SD, PP,
and DD partial wave amplitudes, and $a_4$ contains only the DD partial
wave product.  The angular distributions are normalized for each range
of the center of mass momentum separately, in order to be independent
of the variation of the cross section with the center of mass
momentum.

\begin{figure*}[p]
	\centering
	\includegraphics[width=.95\textwidth]{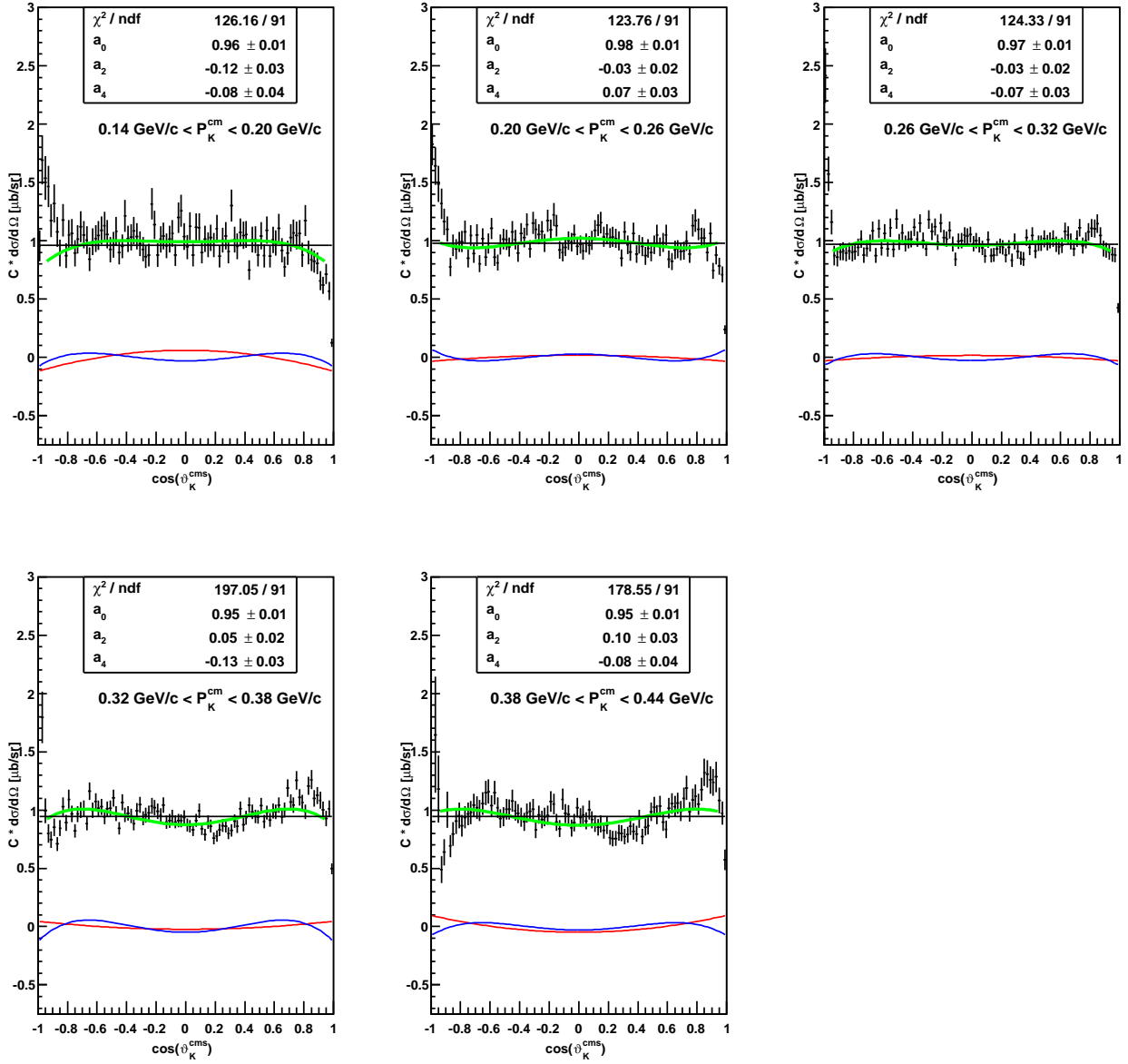}
	\caption{\label{KaonAngularDistributions} Kaon angular distributions for different ranges of the kaon 
        center of mass momentum. The green curve shows the fit result. The contributions
        of the Legendre polynomials  $P_{0}$, $P_{2}$, and $P_{4}$, weighted with the corresponding
        coefficient are plotted in black, red, and blue.  The last figure is missing, as the maximum kaon momentum is less than
        $0.44\,\mathrm{GeV}/\mathrm{c}$.
        }
\end{figure*}

\begin{figure*}[p]
	\centering
	\includegraphics[width=.95\textwidth]{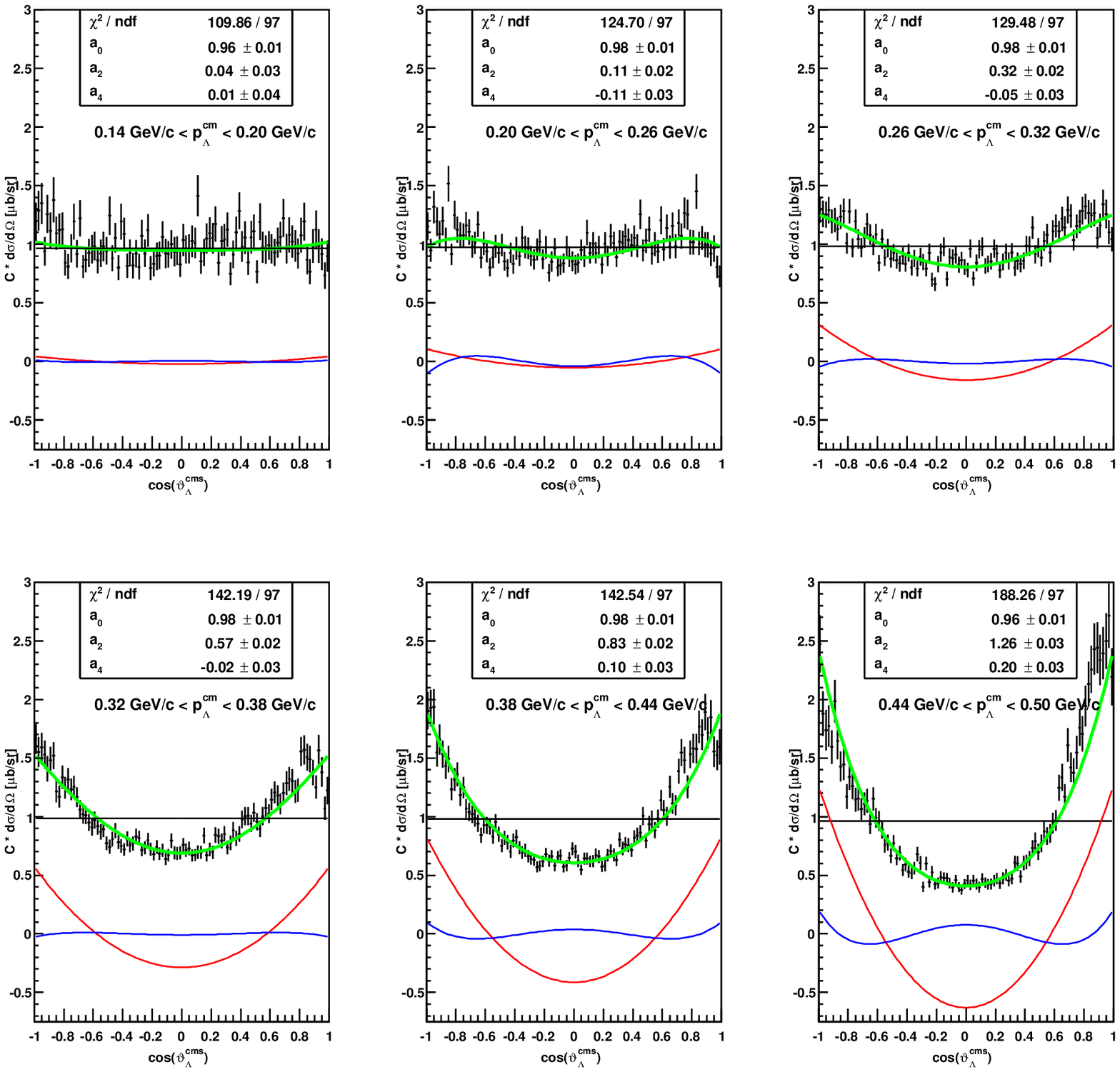}
	\caption{\label{LambdaAngularDistributions}$\mathrm{\Lambda}$ angular distributions for different ranges of the $\mathrm{\Lambda}$ 
        center of mass momentum. The green curve shows the fit result. The contributions
        of the Legendre polynomials  $P_{0}$, $P_{2}$, and $P_{4}$, weighted with the corresponding
        coefficient are plotted in black, red, and blue.
        }
\end{figure*}

The angular distributions are shown in the appendix (fig.~\ref
{ProtonAngularDistributions} for the proton, fig.~\ref
{KaonAngularDistributions} for the kaon and fig.~\ref
{LambdaAngularDistributions} for the $\mathrm{\Lambda}$ particle). In addition
in each figure the Legendre polynomials $P_{0}$, $P_{2}$, and $P_{4}$,
weighted with the corresponding fit coefficients, are plotted and the corresponding
momentum range is indicated.
For the proton and kaon the fit is restricted to the range of
$-0.95 \le \cos\vartheta^{\mathrm{cm}} \le 0.95$. 
The limit in the backward direction takes into account the increase of
background at this region.  The cut in the forward range corresponds
to a limit in the laboratory scattering angle of about \ang{5}, and
takes into account that the proton and kaon have to be detected by the
start-counter, of which the efficiency determination for the very
small angles has a larger error compared to other angles.

\begin{figure}[htbp]
	\begin {center}
	\includegraphics[width=.5\textwidth]{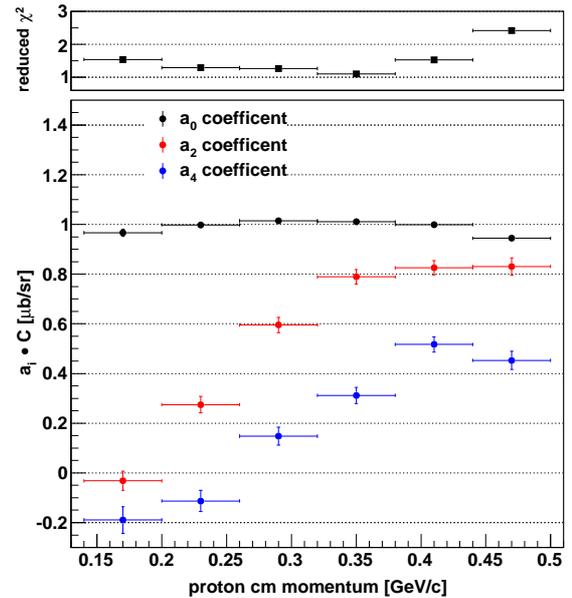}
	\caption{\label{ProtonAngularDistributions-coefficients}The
          dependence of the fit coefficients of the proton angular distribution on the proton center
          of mass momentum is shown.  The upper
        small figure shows the corresponding reduced $\chi^{2}$ of the fit.  }
	\end {center}        
\end{figure}

\begin{figure}[htbp]
	\begin {center}
	\includegraphics[width=.5\textwidth]{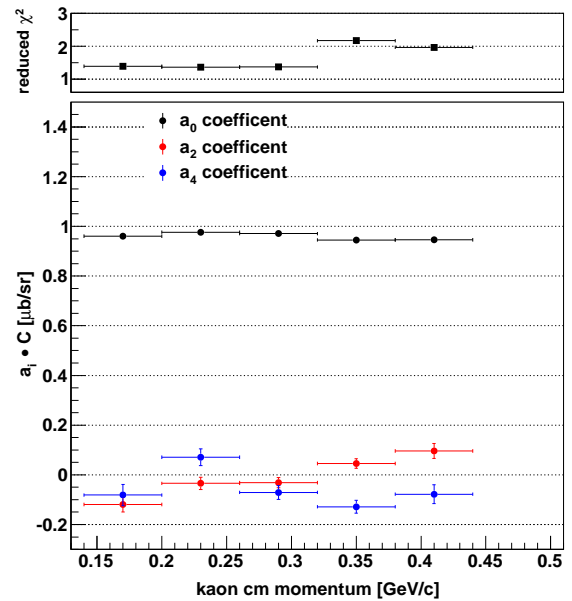}
	\caption{\label{KaonAngularDistributions-coefficients}The
          dependence of the fit coefficients of the kaon angular distribution on the kaon center
          of mass momentum is shown.  The last bin is missing, as the maximum kaon momentum is less than
          $0.44\,\mathrm{GeV}/\mathrm{c}$. The upper
        small figure shows the corresponding reduced $\chi^{2}$ of the fit. }
	\end {center}        
\end{figure}

\begin{figure}[htbp]
	\begin {center}
	\includegraphics[width=.5\textwidth]{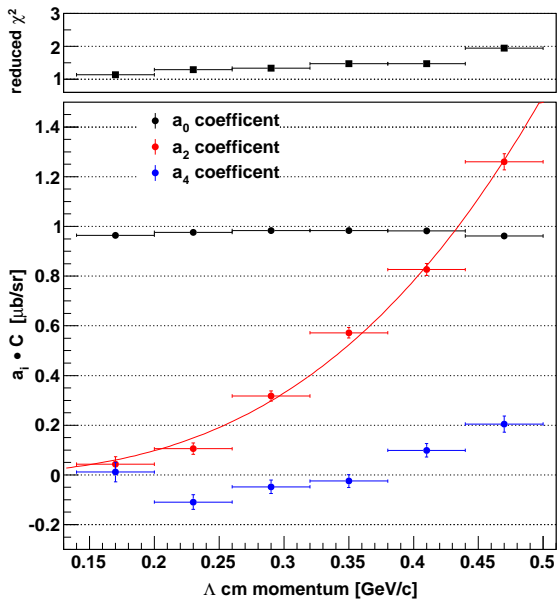}
	\caption{\label{LambdaAngularDistributions-coefficients}The
          dependence of the fit coefficients of the $\mathrm{\Lambda}$
          angular distribution on the $\mathrm{\Lambda}$ center of
          mass momentum is shown.  The upper
        small figure shows the corresponding reduced $\chi^{2}$ of the fit. The distribution of the coefficient
          $a_2$ is fitted with $c\cdot x^{3}$ (continuous red line).
        }
	\end {center}        
\end{figure}

The variations of the coefficients $a_{0}$, $a_{2}$, and $a_{4}$ with
the center of mass momentum are shown in figs. \ref
{ProtonAngularDistributions-coefficients}, \ref
{KaonAngularDistributions-coefficients}, \ref
{LambdaAngularDistributions-coefficients} for p, K, and
$\mathrm{\Lambda}$, respectively.
The coefficient $a_{2}$ of the proton rises nearly linearly from
$0.15\,\mathrm{GeV}/\mathrm{c}$ to $0.35\,\mathrm{GeV}/\mathrm{c}$ and
is constant in the momentum region $0.35 - 0.5\,\mathrm{GeV}/\mathrm{c}$. 
In contrast to this behavior the
coefficient $a_{2}$ of the $\mathrm{\Lambda}$ rises with the third
power of the cm momentum:

\begin{equation*}
a_{2}=1/C\cdot f\cdot (p_{\mathrm{\Lambda}}^{\mathrm{cm}}[\mathrm{GeV}/\mathrm{c}])^{3}
\end{equation*}

with $f = (12.2 \pm 1.4)\,\mu\mathrm{b}/(\mathrm{GeV}/\mathrm{c})^{3}$
(see red line in fig.~\ref{LambdaAngularDistributions-coefficients}).
$C$ is the normalization factor defined in equation \ref
{legendreSum}.

The coefficient $a_{4}$ of the proton starts to be significantly
different from 0 at the cm momentum of $0.3\,\mathrm{GeV}/\mathrm{c}$
and rises up to 
$1/C \cdot 0.6\, [\mu\mathrm{b}/\mathrm{sr}]$ at 
$0.45 - 0.5\,\mathrm{GeV}/\mathrm{c}$. The fit of the $\mathrm{\Lambda}$
distribution gives only for the momentum bin of
$0.5\,\mathrm{GeV}/\mathrm{c}$ a significant $a_{4}$ contribution with
about 1/3 compared to the proton angular distribution.  The kaon
angular distribution can be described for all momentum bins by the
coefficient $a_{0}$ alone, no higher Legendre polynomials are needed
for the fit. The reduced $\chi^2$ of the fit increases only by a
factor of 1.1 to 1.3 by omitting the $P_2$ and $P_4$ polynomials in
the fitting procedure.

The isotropy of the kaon angular distribution, even for the highest
momenta, is in contrast to the behavior of the $\omega$ in the pp
$\rightarrow$ pp$\omega$ reaction \cite {Abdel-Bary2010}. Here the cm
angular distribution of the lighter particle ($\omega$) is strongly
an\-isotropic (for the measurement with a beam momentum of
$3.059\,\mathrm{GeV}/\mathrm{c}$ the maximum cm momentum of the
$\omega$ is with $0.4\,\mathrm{GeV}/\mathrm{c}$ approximately in the
same range of the maximum cm momentum of the kaon). The protons
exhibit a nearly flat angular distribution in contrast to the proton
or $\mathrm{\Lambda}$ of the pK$\mathrm{\Lambda}$ reaction.  A better
comparison is the pp $\rightarrow$ pp$\eta$ reaction, as the mass of
the $\eta$, compared to the one of the $\omega$, is closer to the kaon
mass. Data measured at a beam momentum of
$2.94\,\mathrm{GeV}/\mathrm{c}$, where the $\eta$ has
$0.554\,\mathrm{GeV}/\mathrm{c}$ cm momentum, yield a strongly
an\-isotropic angular distribution. In contrast, the angular distribution is isotropic at
a beam momentum of
$3.67\,\mathrm{GeV}/\mathrm{c}$, corresponding to $\eta$ cm momentum
of $0.774\,\mathrm{GeV}/\mathrm{c}$ \cite {Balestra2004}.  Therefore, there is no clear sign,
whether the cm angular distribution of the lightest particle of a
three particle system is particularly isotropic due to kinematical
effects.

%======================================================================================================================================
\subsection{Invariant mass distributions}\label{detailedInvariantMassDistribution}

%======================================================================================================================================
For better comparability with literature -- especially concerning the
threshold examinations -- the linear invariant mass distributions 
are analyzed. 

The measured invariant mass distribution is given by 

\begin {equation}
  \frac{\mathrm{d}\sigma^{\mathrm{meas}}}{\mathrm{d}m_{\mathrm{ij}}} =\mathrm{PS}(m_{\mathrm{ij}})\cdot
  |\mathrm{M}(m_{\mathrm{ij}})|^{2} \cdot \mathrm{\epsilon}(m_{\mathrm{ij}}) 
\end {equation}

PS denotes the phase-space volume, M($m_{\mathrm{ij}}$) the reaction amplitude, and
$\mathrm{\epsilon}$ is the detector acceptance and reconstruction efficiency,
which are dependent on the invariant mass.  By modeling the reaction
with the Monte Carlo technique the phase-space, the acceptance and reconstruction
efficiency  are calculated:

\begin {equation}
  \frac{\mathrm{d}\sigma^{\mathrm{MC}}}{\mathrm{d}m_{\mathrm{ij}}} = \mathrm{PS}(m_{\mathrm{ij}})\cdot
  \mathrm{\epsilon}(m_{\mathrm{ij}}) 
\end {equation}

The division of the measured distribution by the Monte Carlo
distribution generates a spectrum, which is only dependent on the
reaction amplitude  M($m_{\mathrm{ij}}$) and no assumptions on the
normalization to the phase-space distribution are needed.\footnote{In publications of the
DISTO Collaboration (e.g. \cite {Maggiora2010} ) this distribution is
named ``deviation spectrum''.} In the following these spectra are
analyzed.

\subsubsection {Proton-$\mathrm{\Lambda}$ invariant mass}
The main goal of the examination of this invariant mass distribution
is the analysis of the behavior at the N$\mathrm{\Sigma}$
threshold. In the strict sense there are two thresholds:
n$\mathrm{\Sigma}^+$ at $2.129\,\mathrm{GeV}/\mathrm{c}^2$ and
p$\mathrm{\Sigma}^0$ at $2.131\,\mathrm{GeV}/\mathrm{c}^2$.  This mass
difference is about $1\sigma$ of the invariant mass resolution and not
resolved in the data. In the following calculations the mass of the
lower threshold is used, as this channel dominates the cusp effect
\cite {Haidenbauer2015}. For convenience the threshold is named
N$\mathrm{\Sigma}$ threshold.

In order to examine the  N$\mathrm{\Sigma}$ threshold we assume as a working hypothesis
that the final state interaction (FSI), the N$\mathrm{\Sigma}$ threshold (TH), and the
reflections of  N* resonances (RF) are independent and interference
terms are negligible:
\begin {equation}\label {eq:invMassSum}
  \frac{\mathrm{d}\sigma^{\mathrm{meas}}}{\mathrm{d}m_{\mathrm{p\Lambda}}}/\frac{\mathrm{d}\sigma^{\mathrm{MC}}}{\mathrm{d}m_{\mathrm{p\Lambda}}}
  = \mathrm{FSI}(m_{\mathrm{p\Lambda}}) + \mathrm{TH}(m_{\mathrm{p\Lambda}}) +\mathrm{RF}(m_{\mathrm{p\Lambda}})
\end {equation}

Of course this is a simplification, as the reflections of the N* resonances
can interfere with the final state interaction and the threshold effect. 
The final state interaction, the reflection of N* resonances and the threshold effect are
described with functions with free parameters. The parameters are
determined by fitting the invariant mass distribution with the sum of
these functions.

The final state interaction is treated like the other terms as an additive term 
in order to incorporate that the FSI has strong influence
at low p$\mathrm{\Lambda}$ invariant mass and this influence   is negligible
towards larger invariant masses.\footnote{A similar behavior 
  is measured for the pp final state in $pp \rightarrow
  pp\eta$ at an excess energy of $\SI{41}{\mega\electronvolt}$ \cite {Abdel-Bary2003a} and in
  $np \rightarrow pp\pi^-$ at excess energies between 405 and $\SI{465}{\mega\electronvolt}$
\cite {Daum2002}.}
This is in
contrast to the  ansatz, where the reaction amplitude is
factorized with the FSI amplitude (see for instance \cite
{Sibirtsev2006} and references therein). 

%======================================================================================================================================

The final state interaction is parameterized with 

\begin {equation} \label{eq:fsi1}
  \mathrm{FSI}(m_{\mathrm{p\Lambda}}) = \mathrm{exp}(c_0 + \frac{c_1}{ m_{\mathrm{p\Lambda}}^{2}-c_2}) 
\end {equation}

The effective scattering length can be expressed by two of the three
coefficients of equation \ref{eq:fsi1}:

\begin {equation} \label{eq:fsi2}
 a_{\mathrm{eff}} = -1/2\, \hbar c\, c_1 \sqrt{\frac{m_{\mathrm{\min}}^{2}}{m_{\mathrm{p}}
      m_{\mathrm{\Lambda}}} \cdot \frac{m_{\mathrm{\max}}^2 -m_{\mathrm{\min}}^2}{ (m_{\mathrm{\max}}^2 -c_2)
      \cdot (m_{\mathrm{\min}}^2 -c_2)^3}}
\end {equation}

$m_{\mathrm{min}}$ and $m_{\mathrm{max}}$ denote the range of the invariant mass
$m_{\mathrm{p\Lambda}}$ in which this formula is applied to determine the
scattering length.  Equations \ref {eq:fsi1} and \ref{eq:fsi2} are
given in \cite{Hauenstein2014}, they are derived in
\cite{Gasparyan2005}. Equation \ref {eq:fsi2} is applied to express
the coefficient $c_1$ with the effective scattering length. The value
of $a_{\mathrm{eff}} = \SI[parse-numbers=false]{1.233}{\femto\metre}$ is taken, which is deduced from  data
measured at a beam momentum of $2.7\,\mathrm{GeV}/\mathrm{c}$ \cite {Hauenstein2014}. 
At this beam momentum no disturbing effects of the N$\Sigma$ threshold are present.
The influence of the N* resonances were tested by evaluating  different regions of
the Dalitz plot, they give limits of the measured effective scattering length between
-1.4 and -1.2 fm \cite {Hauenstein2014}, at the beam momentum of 2.95 these limits were found to be larger with
-0.9 and -2.1 fm \cite {Roeder2013}.
With this procedure the final state interaction
at $2.95\,\mathrm{GeV}/\mathrm{c}$ is expressed with two parameters $c_0$ and $c_2$.
%======================================================================================================================================

%\subsubsubsection{Background}
The nucleon resonances N*(1650), N*(1710), 
N*(1720) contribute to deviations from the phase-space in the Dalitz
plot \cite {AbdEl-Samad2010}.  As these resonances decay into
$\mathrm{K^+\Lambda}$ their influence can be seen directly  in the invariant
mass distribution of $m_{\mathrm{K^+\Lambda}}$ and as reflections in the
invariant mass distribution of $m_{\mathrm{p\Lambda}}$. As their widths are in
the range of $50 - 400\,\mathrm{MeV}$ \cite {Olive2014} the reflections are slowly
varying with $m_{\mathrm{p\Lambda}}$ compared to the FSI and the threshold
effect, whose structures are smaller than $20\,\mathrm{MeV}/\mathrm{c}^2$. Therefore, these
reflections are parameterized as a polynomial of  second order. The
main purpose of this parametrization is to quantify the background
beneath the cusp effect, and not to draw conclusions on these
resonances. For simplicity this part is named ``reflections'' in the further 
description.  
%======================================================================================================================================

As the threshold effect is expected to be a coupled channel effect, we apply the
Flatt\'{e} formalism \cite{Flatte1976}. This formalism was developed to
describe the  $\eta \pi$ ($ \pi \pi$)  interaction at the two kaon
threshold. It is  assumed that the $a_0(980)$ ($f_0(980)$) resonance changes its
width due to  the opening channel of KK thus inducing a cusp structure.
This formalism is adapted to the p$\mathrm{\Lambda}$ channel by
assuming a virtual state of p$\mathrm{\Lambda}$, which decays into  p$\mathrm{\Lambda}$
and into N$\mathrm{\Sigma}$, thus causing a cusp effect \cite {AbdEl-Samad2013}.

\begin {equation} \label{eq:cusp1}
  \mathrm{TH}(m_{\mathrm{p\Lambda}})=
C|\frac{m_{\mathrm{r}}\sqrt{\Gamma_{0}\cdot\Gamma_{ \mathrm{\Lambda p}}}}
{m_{\mathrm{r}}^2-m_{ \mathrm{\Lambda p}}^2-im_{\mathrm{r}}(\Gamma_{ \mathrm{\Lambda p}}+\Gamma_{ \mathrm{\Sigma N}})}|^2
\end {equation}

$m_{\mathrm{r}}$ and $\Gamma_{0}$ are the mass and the total width of a virtual
p$\mathrm{\Lambda}$ state. $C$ is a normalization parameter. As it cannot be
separated from the unknown total width $\Gamma_0$, the product $C
\!\cdot\! \Gamma_0$ is treated as one parameter for the fitting
procedure. $\Gamma_{ \mathrm{\Lambda p}}$ and $\Gamma_{ \mathrm{\Sigma N}}$ are the partial
decay widths of the virtual state, which are given by:

\begin {equation} \label{eq:cusp2}
  \Gamma_{\mathrm{\Lambda p}} = g_{\mathrm{\Lambda p}} \cdot q_{\mathrm{\Lambda p}} \quad
  \textrm{and} \quad \Gamma_{\mathrm{\Sigma N}} = g_{\mathrm{\Sigma N}} \cdot q_{\mathrm{\Sigma N}} 
\end {equation}

$ g_{\mathrm{\Lambda p}}$ and  $g_{\mathrm{\Sigma N}}$ are the coupling constants of the
two channels to the virtual   p$\mathrm{\Lambda}$ state and  $q_{\mathrm{\Lambda p}}$
and  $q_{\mathrm{\Sigma N}}$ are the cm momenta in the corresponding two-body
subsystems.  In dependence on the invariant mass $m_{\mathrm{p\Lambda}}$,
$q_{\mathrm{\Sigma N}}$ is imaginary below the N$\mathrm{\Sigma}$ threshold:

\begin {equation} \label{eq:cusp3}
  q_{\mathrm{\Sigma N}}=i\frac{\sqrt{((m_{\mathrm{\Sigma^+}}+m_{\mathrm{n}})^2-m_{\mathrm{p\Lambda}}^2)\cdot(m_{\mathrm{p\Lambda}}^2-(m_{\mathrm{\Sigma^+}}-m_{\mathrm{n}})^2)}}{2m_{\mathrm{p\Lambda}}}  
\end {equation}

and real above the  N$\mathrm{\Sigma}$ threshold:

\begin {equation} \label{eq:cusp4}
  q_{\mathrm{\Sigma N}}=\frac{\sqrt{(m_{\mathrm{p\Lambda}}^2-(m_{\mathrm{\Sigma^+}}+m_{\mathrm{n}})^2)\cdot(m_{\mathrm{p\Lambda}}^2-(m_{\mathrm{\Sigma^+}}-m_{\mathrm{n}})^2)}}{2m_{\mathrm{p\Lambda}}}  
\end {equation}

The cm momentum $q_{\mathrm{\Lambda p}}$ is always real and given by
exchanging the $\mathrm{\Sigma^+}$ and neutron mass
($m_{\mathrm{\Sigma^+}}$, $m_{\mathrm{n}}$) with the
$\mathrm{\Lambda}$ and proton mass ($m_{\mathrm{\Lambda}}$,
$m_{\mathrm{p}}$) in equation \ref {eq:cusp4}. A detailed examination
of Flatt\'{e}-like distributions \cite {Baru2005} reveals that the
coupling constants and the mass of the virtual state show a scaling
behavior.
%die aussage ist genau richtig nur für die Resonanz in der Nähe der Schwelle
 Therefore, the coupling constant $g_{\mathrm{\Lambda p}}$ is fixed to
 the arbitrarily chosen value of 1, and only the remaining parameters
 $C\!\Gamma_0$, $g_{\mathrm{\Sigma N}}$, and $m_{\mathrm{r}}$ are
 determined together with the parameters for the reflections and FSI by
 a fit procedure.

\begin{figure}[htbp]
	\centering
	\includegraphics[width=.5\textwidth]{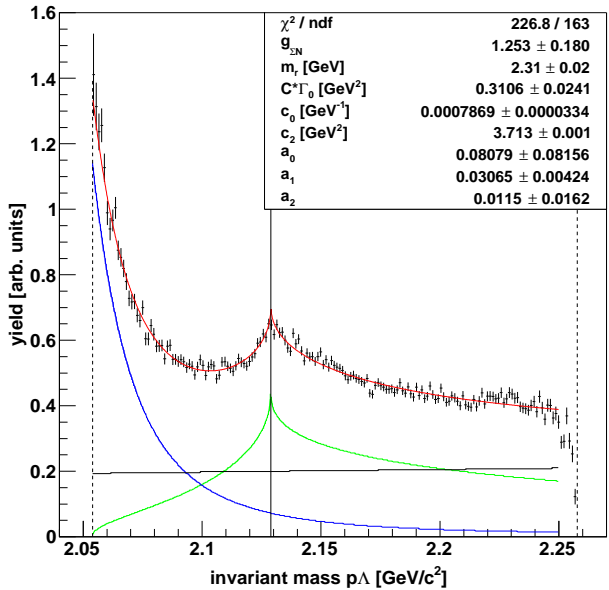}
        %aus flatte-hel/work.C erstellt
	\caption{\label{mpl-gesamt-Fit}The whole data set divided by
          the Monte Carlo data is shown as dots with error bars. The
          fit result is given by the red line, it is composed of the
          FSI part (blue line), the  part of the reflections (black line), and
          the Flatt\'{e}-distribution (green line). The N$\mathrm{\Sigma}$
          threshold is marked with the vertical black line. The
          kinematic limits are indicated by two  dashed lines.}
\end{figure}

%======================================================================================================================================
The measured invariant mass distribution -- divided by the Monte Carlo
distribution -- fitted with the function as described in equation
\ref {eq:invMassSum} is shown in fig.~\ref {mpl-gesamt-Fit}. The fit starts at
the kinematic limit of $m_{\mathrm{p}} + m_{\mathrm{\Lambda}}$ and is limited to $2.250\,\mathrm{GeV}/\mathrm{c}^2$, 
which is $8\,\mathrm{MeV}/\mathrm{c}^2$ below the maximum possible kinematical value.
This limit is chosen as the data and Monte Carlo distributions have above this value
a sharp decrease to zero, therefore, the uncertainties of the ratios are large.   

The structures of the data are well described by the fit. The sharp
rising leading edge, the peak at the threshold, and the slow
decreasing above the threshold is reproduced by the Flatt\'{e}
formalism. The mass of the virtual p$\mathrm{\Lambda}$ state is
$2.31\,\mathrm{GeV}/\mathrm{c}^2$.  However, this value is depending
on the arbitrarily chosen value of the p$\mathrm{\Lambda}$ coupling
constant to this virtual state. Changing this coupling constant for
instance from 1 to 0.5 produces a  fit result for the mass of the virtual
p$\mathrm{\Lambda}$ state of
$2.20\,\mathrm{GeV}/\mathrm{c}^2$.  As remarked in \cite {Flatte1976},
the mass of the virtual state has to be larger than the threshold in
order to generate a sharp cusp shape. For masses below the threshold,
the distribution varies more smoothly, and is more similar to a
Breit-Wigner distribution. The coupling constant of the virtual state
to the N$\mathrm{\Sigma}$ channel ($g_{\mathrm{\Sigma N}}$ = 1.253 $\pm$ 0.189)  is larger than the coupling
constant to the p$\mathrm{\Lambda}$ ($g_{\mathrm{\Lambda p}}$ = 1, fixed value) channel. This is in contrast to
the previous analysis \cite{AbdEl-Samad2013}, which applied a
different mass of the virtual state. The first attempt to describe the
threshold behavior of the p$\mathrm{\Lambda}$ invariant mass
distribution with the Flatt\'{e} formula was performed in \cite
{Braun1977}, where a measurement of K$^{-}$d $\rightarrow$
p$\mathrm{\Lambda}\mathrm{\pi}^{-}$ is described. Here the best fit
result is obtained with a mass of $2.20\,\mathrm{GeV}/\mathrm{c}^2$ as input parameter.
%======================================================================================================================================

\begin{figure}[htbp]
	\centering
	\includegraphics[width=.5\textwidth]{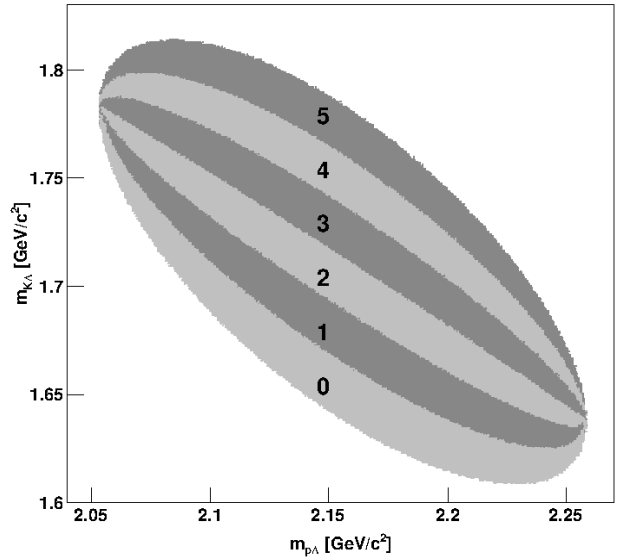}
	\caption{\label{helicity-slices}The division of the Dalitz
          plot into six slices by selecting ranges of  the helicity angle between
          proton and kaon in the p$\mathrm{\Lambda}$ subsystem is shown. If the Dalitz
          plot is filled according to the phase-space, the
          projection of the slices $i =0 .. 5$ onto the invariant mass
          p$\mathrm{\Lambda}$ results in identical distributions.   }
\end{figure}

As can be seen in the Dalitz plot of fig.~\ref{dalitz-pl-kl}, the density
in the N$\mathrm{\Sigma}$ threshold region, is strongly decreasing with increasing $m_{\mathrm{K\Lambda}}$.
 In order to examine, whether this variation is
due to the cusp effect or due to the N* contributions, the
Dalitz plot is divided into 6 slices, by applying cuts on the 
helicity angle between the proton and the kaon in the p$\mathrm{\Lambda}$ subsystem 
(s. fig.~\ref {helicity-slices}). The limits of 
each slice are given by 

\begin {equation}
 -1+i\cdot1/3 < \cos \vartheta_{\mathrm{pK}}^{\mathrm{Rp\Lambda}} <-1+(i+1)\cdot1/3 
\end {equation}

with $i = 0 .. 5$. 
The projection of these slices onto the p$\mathrm{\Lambda}$ invariant mass results in 
identical distributions if the Dalitz plot is filled according to phase-space.  

\begin{figure}[htbp]
	\centering
	\includegraphics[width=.5\textwidth]{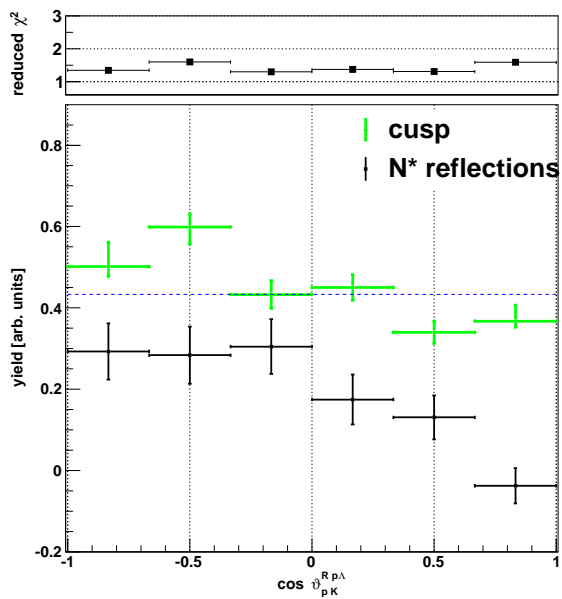}
	\caption{\label{helicity-slices-flatte-bg} The
          height of the cusp (green crosses) and the amount of the
          reflections at the cusp position (black crosses) for the
          different regions of the Dalitz plot as shown in
          fig.~\ref{helicity-slices}. The blue dashed line shows the
          fit result of the height of the cusp of the total data set.
          As the errors of the fit parameter  $C\!\Gamma_0$ are found
          to be asymmetric, the errors of the MIGRAD routine of
          the ROOT \cite {Brun2003} program are shown.  The upper
        small figure shows the corresponding reduced $\chi^{2}$ of the fit.
        }
\end{figure}

For each slice the invariant mass p$\mathrm{\Lambda}$ is fitted with the same
function as applied for the whole data set. As the FSI distribution is
expected to be independent of the helicity angle, the parameters
$c_{0}$, $c_{2}$ are fixed to the values which are obtained by the fit
of the whole data set. In addition it is assumed that the cusp does
not change its shape, therefore, the parameters $m_{\mathrm{r}}$, and
$g_{\mathrm{N\Sigma}}$ are fixed to the values obtained by the whole data fit.
$C\!\Gamma_0$ and the coefficients of the reflections are determined
for each slice with the same fit procedure.  The height of the cusp is
proportional to $C\!\Gamma_0$. With $m_{\mathrm{r}} = 2.31\,\mathrm{GeV}/\mathrm{c}^2$ the
proportional factor is 1.39.  As the shape of the cusp is constant,
$C\!\Gamma_0$ is in addition proportional to the area beneath the cusp
described by the Flatt\'{e} function.  The amount of the reflections beneath the cusp
is determined as the value of the second order polynomial at the cusp
position ($m_{\mathrm{p\Lambda}} = 2.129\,\mathrm{GeV}/\mathrm{c}^2$).  The individual
distributions of the slices and their fit results are shown in the
appendix (fig.~\ref{helicity-slices-fits}). The variation of the
height of the Flatt\'{e} distribution and the amount of the reflections at the
threshold is shown in fig.~\ref {helicity-slices-flatte-bg}.
While the height of the Flatt\'{e} distribution varies only slowly with
the helicity angle, the  part of the reflections is nearly constant for the
first three slices and then drops from 0.3 to 0 in the last three
slices. This indicates that the large intensity variation in the
Dalitz  plot along the cusp line is caused by the background generated 
by the N$^*$ resonances.

The assumption, that the cusp does not change the shape, seems to be justified 
by the fit results. It can be seen in fig.~\ref{helicity-slices-fits}, that
the cusp is well described by the fit function, having the same parameters,
which affect the cusp shape. The quality of the fit is nearly the same for
each slice, all reduced $\chi^2$  are in the range of $1.4 \pm 0.2$.  
%======================================================================================================================================

\begin{figure*}[p]
	\centering
	\includegraphics[width=.95\textwidth]{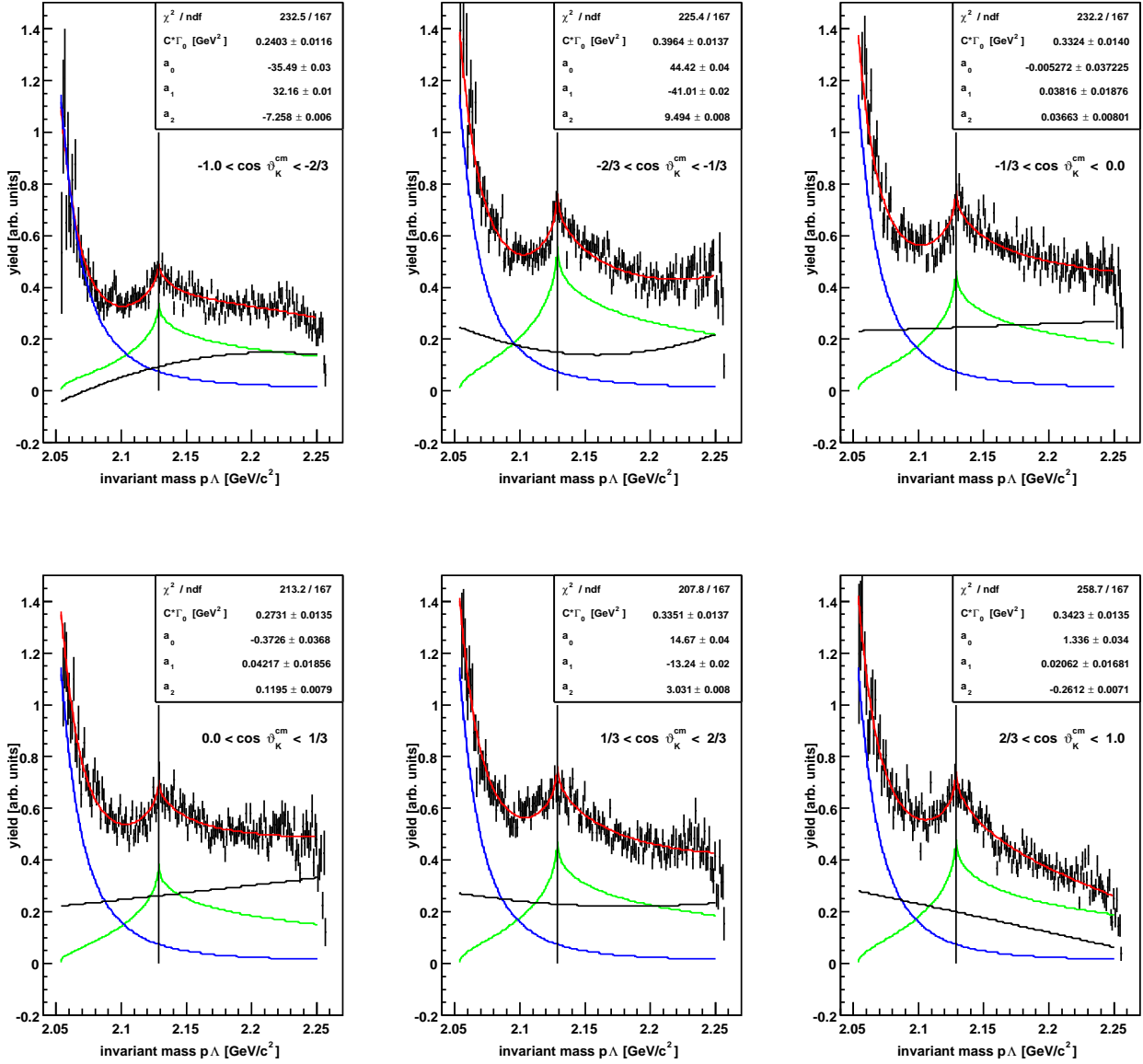}
	\caption{\label{kcm-slices-fits} The p$\mathrm{\Lambda}$ invariant
          mass distributions are shown for different slices 
          of the kaon cm scattering angle as given in the individual plots.
          The spectra are normalized to
          1/6 of the total data set which is shown in
          fig.~\ref{mpl-gesamt-Fit}.  The data are shown as points
          with error bars, the red line is the total fit.  The green
          line is the contribution of the Flatt\'{e} distribution, where
          only the height ($C\!\Gamma_0$) is varied.
          The blue line is the  FSI part with all parameters
          fixed. The black line is the  part of the reflections with freely
          varying parameters of the second order polynomial. }
\end{figure*}

The angular distribution of the cusp effect reflects the angular
distribution of the $\mathrm{p\Lambda}$ system, which stems from the
N$\mathrm{\Sigma}$ system at the threshold.  The angular momenta
between p and $\mathrm{\Lambda}$ inside the $\mathrm{p\Lambda}$ system
are described by the Gottfried-Jackson angular distribution
($\cos\vartheta_{\mathrm{pb}}^{\mathrm{Rp\Lambda}}$). As the
$\mathrm{p\Lambda}$ system has the same spin-parity as the N$\Sigma$
system at the threshold ($J^{P} = 0^{+},1^{+}$), it can only be in
relative S or D waves \cite{AbdEl-Samad2013}.  The angular momentum of
the cusp can be examined with the kaon angular distribution in the cm
system, which is one of the two particle subsystem K-(N$\Sigma$) and
thus the kaon angular distribution is the mirrored one of
(N$\Sigma$). This angular distribution has to be symmetric to
$\cos\vartheta_{\mathrm{K}}^{\mathrm{cm}} = 0$.  Both angular
distributions $\cos\vartheta_{\mathrm{K}}^{\mathrm{cm}}$ and
$\cos\vartheta_{\mathrm{pb}}^{\mathrm{R p\Lambda}}$ of the height of
the Flatt\'{e} distribution and  the amount of the reflections beneath the cusp are
generated with the slice method as described in the previous
section. In this case $\cos\vartheta_{\mathrm{K}}^{\mathrm{cm}}$ and
$\cos\vartheta_{\mathrm{pb}}^{\mathrm{R p\Lambda}}$ are divided into 6
bins.  The fit results for the
$\cos\vartheta_{\mathrm{K}}^{\mathrm{cm}}$ slices are shown in
fig.~\ref{kcm-slices-fits} and for the
$\cos\vartheta_{\mathrm{pb}}^{\mathrm{R p\Lambda}}$ slice in fig.
\ref{gfj-slices-fits} in the appendix.

\begin{figure}[htbp]
	\centering
	\resizebox{1.\columnwidth}{!}{\includegraphics{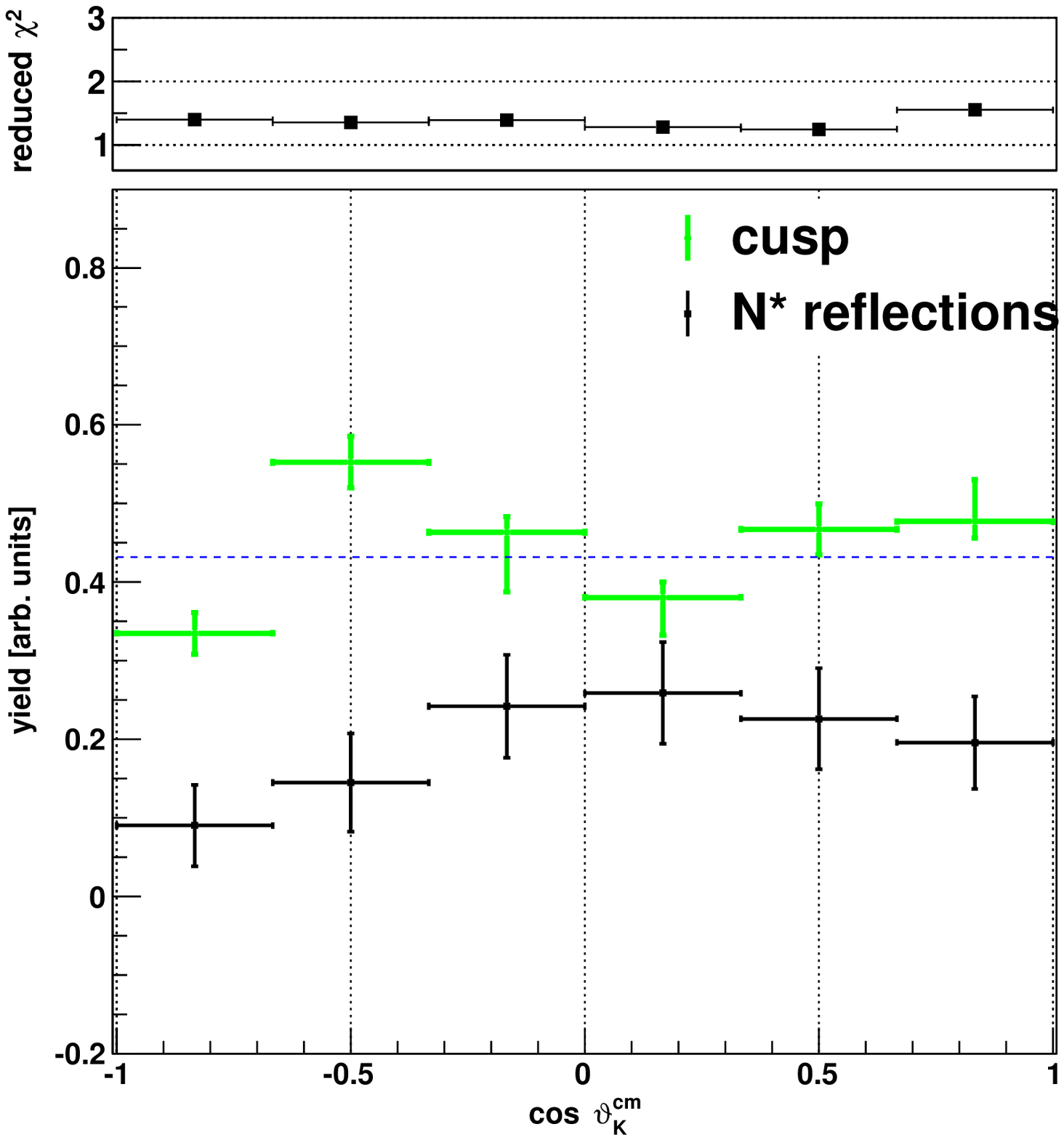}}
	\caption{\label{kcm-slices-flatte-bg} The height of the cusp
          (green crosses) and the amount of the reflections at the cusp
          position (black crosses) for the different regions of
          $\cos\vartheta_{\mathrm{K}}^{\mathrm{cm}}$. The blue dashed
          line shows the fit result of the height of the cusp of the
          total data set.  As the errors of the fit parameter
          $C\!\Gamma_0$ are found to be asymmetric, the errors of the
          MIGRAD routine of the ROOT \cite {Brun2003} program are
          shown.  The upper
        small figure shows the corresponding reduced $\chi^{2}$ of the fit.
        }
\end{figure}

\begin{figure}[htbp]
	\centering
	\includegraphics[width=.5\textwidth]{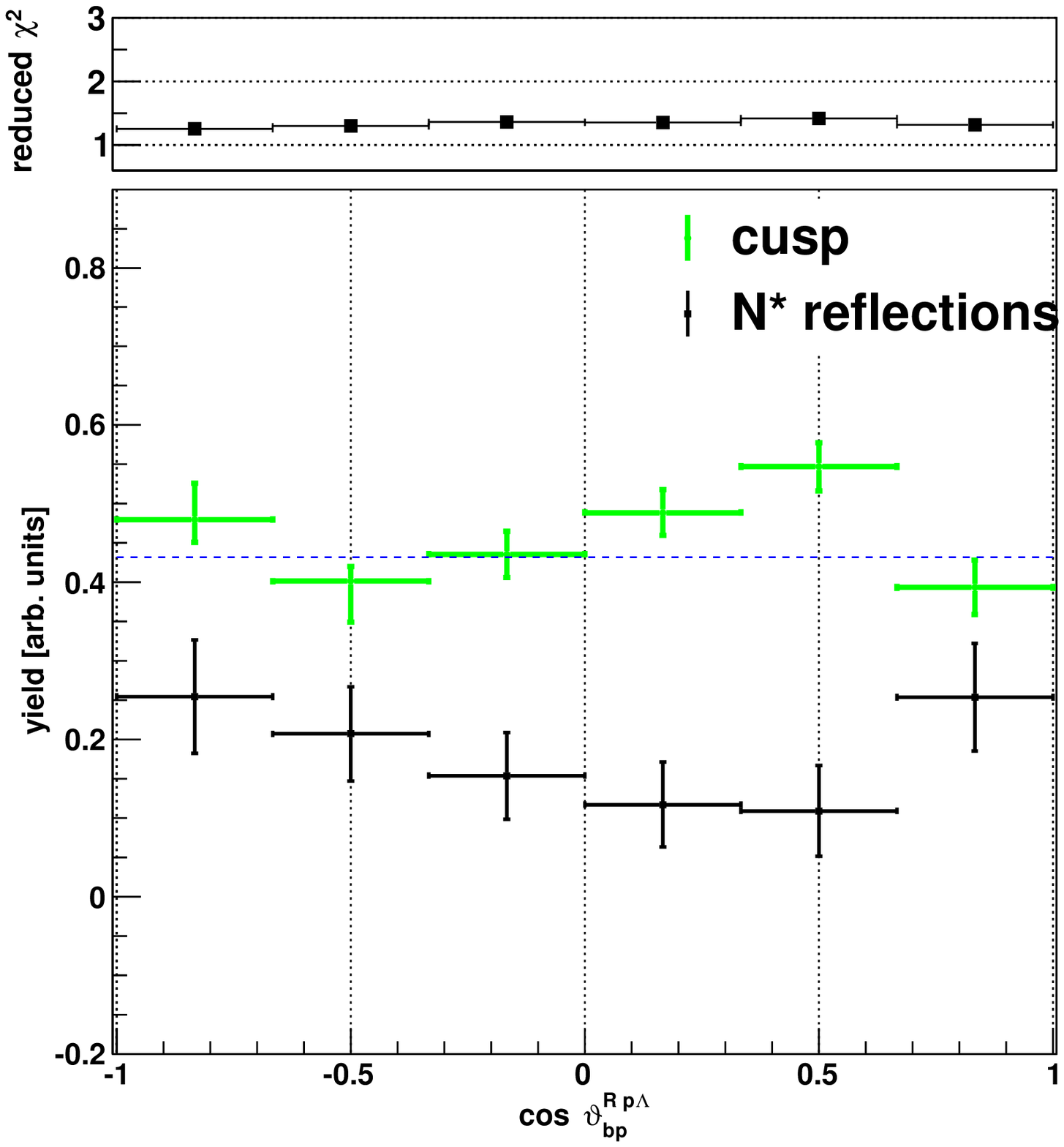}
	\caption{\label{gfj-slices-flatte-bg} The height of the cusp
          (green crosses) and the amount of the reflections at the cusp
          position (black crosses) for the different regions of
          $\cos\vartheta_{\mathrm{bp}}^{\mathrm{R p\Lambda}}$. The
          blue dashed line shows the fit result of the height of the
          cusp of the total data set.  As the errors of the fit
          parameter $C\!\Gamma_0$ are found to be asymmetric, the
          errors of the MIGRAD routine of the ROOT \cite {Brun2003}
          program are shown.  The upper
        small figure shows the corresponding reduced $\chi^{2}$ of the fit.
        }
\end{figure}

The variation of the height of the Flatt\'{e}
distribution and the amount of the reflections at the threshold in dependence on the
different slices is shown in figs. \ref {kcm-slices-flatte-bg}, \ref {gfj-slices-flatte-bg}.


No systematic deviations of the cusp height distributions from isotropy, which
could be a hint for P or D waves, can be seen. This is a confirmation
of the findings in \cite{AbdEl-Samad2013}, which are based on a
different data set of COSY-TOF and on a different analysis method. The
shape of the cusp does not change with the examined angles, as can be
seen from the fits of the individual slices. The reduced $\chi^2$ of
all fits does not deviate from the mean value of 1.36 by more than
0.2.
%======================================================================================================================================
\subsubsection {Kaon-$\mathrm{\Lambda}$ invariant mass}\label{detailedGJ}
Deviations of the K$\mathrm{\Lambda}$ invariant mass from phase-space can be caused by a K$\mathrm{\Sigma}$ threshold effect or by nucleon
resonances, which decay into the  K$\mathrm{\Lambda}$ channel. In addition
reflections of the FSI and the  N$\mathrm{\Sigma}$ cusp in the  p$\mathrm{\Lambda}$
invariant mass distributions contribute to the deviations. 

\begin{figure}[htbp]
	\centering
	\includegraphics[width=.5\textwidth]{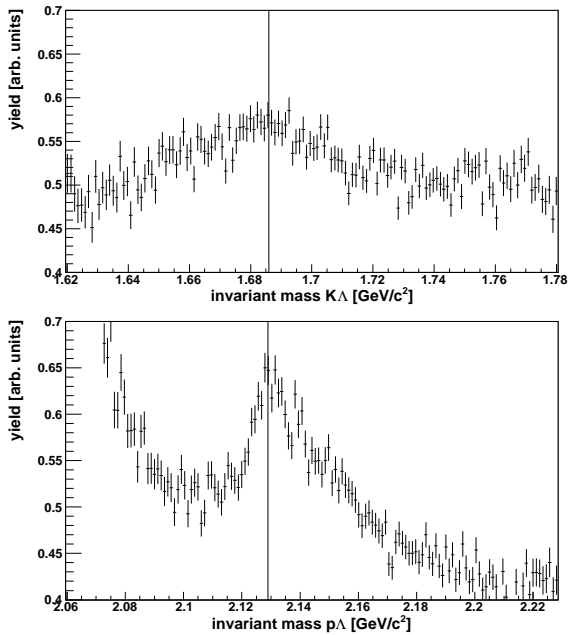}
	\caption{\label{thresholdcomparison}
          The K$\mathrm{\Sigma}$ threshold region of the
          K$\mathrm{\Lambda}$ invariant mass is shown (top) in
          comparison to the N$\mathrm{\Sigma}$ threshold region of the
          p$\mathrm{\Lambda}$ invariant mass (bottom).  The thresholds
          are indicated by the vertical lines. In contrast to the
          strong and sharp enhancement at the N$\mathrm{\Sigma}$
          threshold in the lower figure, no sharp structure is seen at the
          K$\mathrm{\Sigma}$ threshold in the upper figure.}

\end{figure}

At the K$\mathrm{\Sigma}$ threshold no cusp-like structure is
detected, as shown in fig.~\ref {thresholdcomparison}. Here the
K$\mathrm{\Sigma}$ threshold region is compared to the
N$\mathrm{\Sigma}$ threshold region. The effect of the
p$\mathrm{\Sigma}$ cusp is estimated to be
$\SI[parse-numbers=false]{(15 \pm 5)}{\percent}$ of the total cross
section, which corresponds to $\SI[parse-numbers=false]{(3 \pm
  1)}{\micro\barn}$.
%höhe des cusps :.04 , fehlerbalken .003 -> 1/10 We estimate the
From fig.~\ref{thresholdcomparison} it can be seen that the height of
the N$\mathrm{\Sigma}$ cusp compared to the error bars of the
measurement is about 10/1, therefore, the upper limit of a possible
K$\mathrm{\Sigma}$ cusp - if it has a similar shape as the
N$\mathrm{\Sigma}$ cusp - is in the order of $\SI{0.3}{\micro\barn}$.

The contribution of N$^*$ resonances decaying into K$\mathrm{\Lambda}$
will vary with the distance of the invariant mass K$\mathrm{\Lambda}$
to the resonance mass. As N$^*$ resonances affect the composition of
the angular momenta, this composition is tested for different ranges
of the invariant mass K$\mathrm{\Lambda}$. The angular distribution
for the K$\mathrm{\Lambda}$ rest system given by the cosine of the
Gottfried-Jackson angle
$\vartheta^{\mathrm{RK\Lambda}}_{\mathrm{b\Lambda}}$ is generated for
six ranges of the K$\mathrm{\Lambda}$ invariant mass. The
distributions and the corresponding ranges are shown in the appendix
in fig.~\ref {invKL-slices-fits}.  The angular distributions are
fitted with a formula describing the differential cross section as a
sum of Legendre polynomials by neglecting the interference terms. As
the angular distributions do not have to be symmetric to
$\cos\vartheta =0$ -- as it is the case of the center of mass system
-- the odd terms of the Legendre polynomials have to be included.

\begin {equation} \label{legendrePolynomial} 
 \frac{\mathrm{d}\sigma}{\mathrm{d}cos\vartheta^{\mathrm{RK\Lambda}}_{\mathrm{b\Lambda}}}= C\cdot \sum_{i=0}^{n} (a_{i}P_{i}(\cos\vartheta^{\mathrm{RK\Lambda}}_{\mathrm{b\Lambda}}))
\end {equation}

\begin{figure}[htbp]
	\centering
	\includegraphics[width=.5\textwidth]{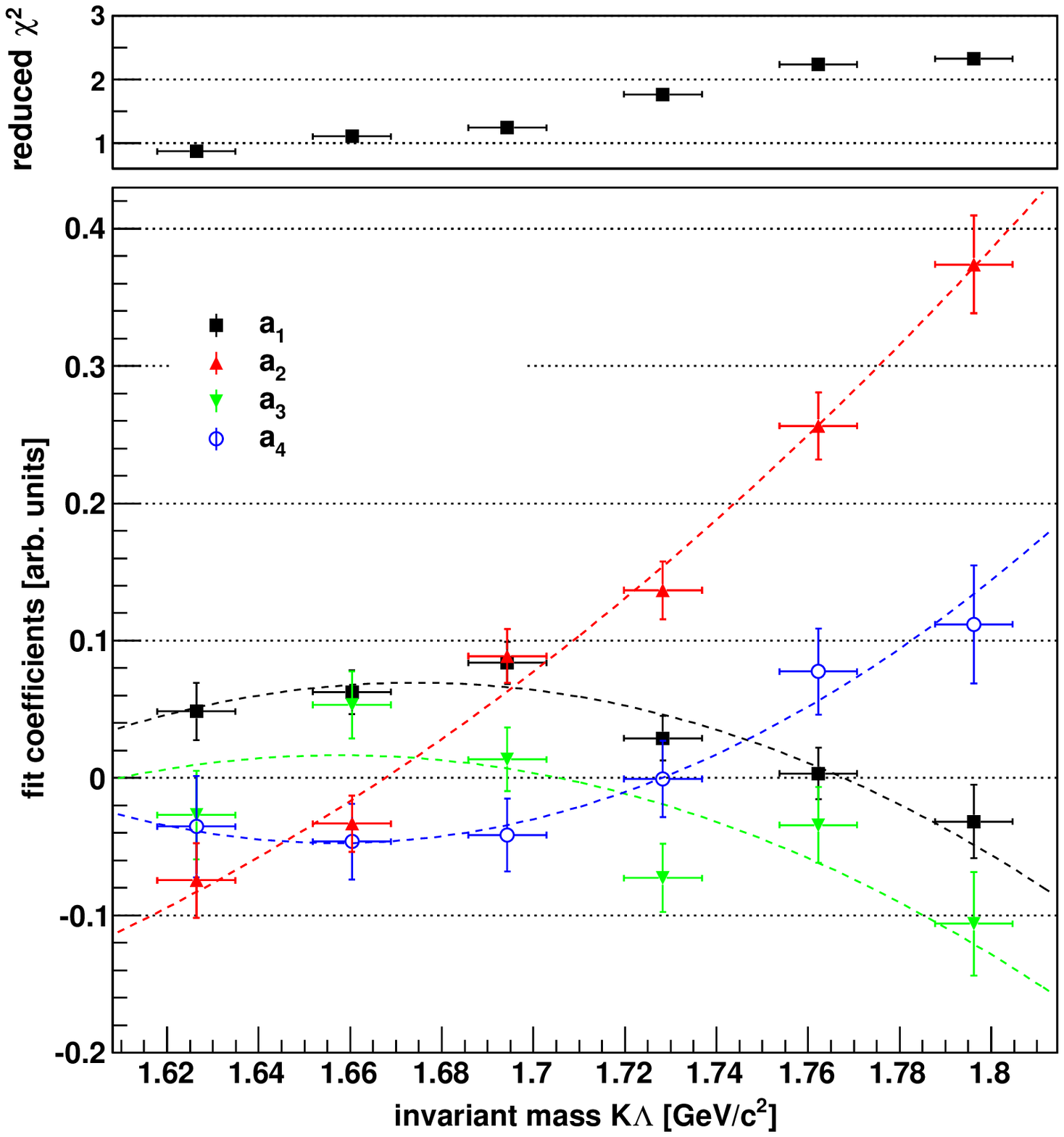}
	\caption{\label{invKL-coefficients}
        Fit results of the coefficients of equation \ref{legendrePolynomial} obtained
        from the GJ angular distributions $\cos\vartheta^{\mathrm{RK\Lambda}}_{\mathrm{b\Lambda}}$ for six
        bins of the K$\mathrm{\Lambda}$ invariant mass with equal widths (lower figure). The coefficient $a_0$
        is close to 1, it is given in fig.~\ref{invKL-slices-fits}. The upper
        figure shows the corresponding reduced $\chi^{2}$ of the fit. The dashed lines
        are generated with a second degree polynomial fit and have the only
        purpose to guide the eye.}
\end{figure}

$C$ is a normalization constant, for these fits the highest order
$n=4$ is chosen.  The dependence of the Legendre coefficients $a_{i}$
on the K$\mathrm{\Lambda}$ invariant mass slices is shown in
fig.~\ref{invKL-coefficients}.

\begin{figure}[htbp]
	\centering
	\includegraphics[width=.5\textwidth]{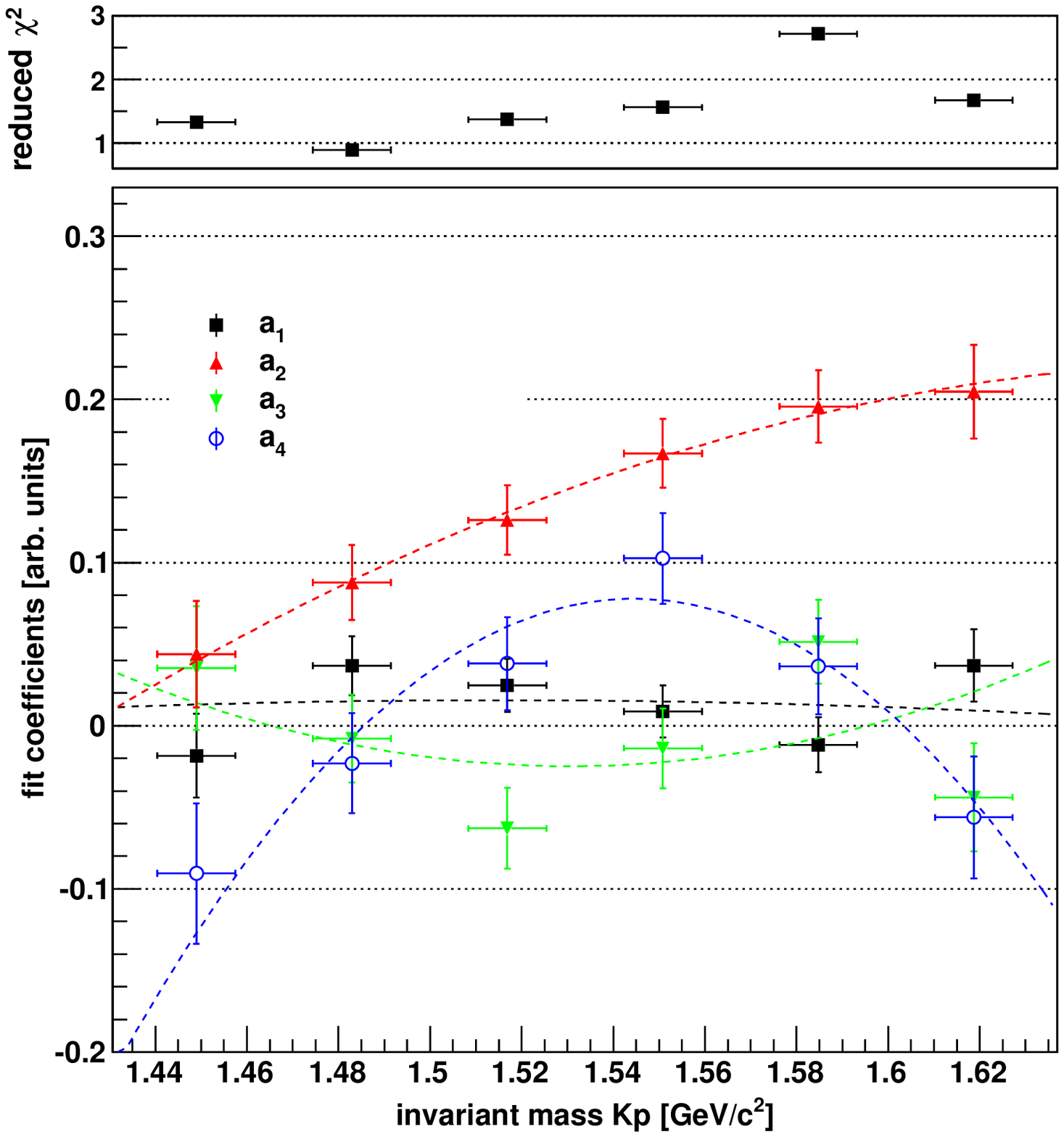}
	\caption{\label{invKp-coefficients}
          Fit results of the coefficients of equation \ref{legendrePolynomial} obtained
        from the GJ angular distributions $\cos\vartheta^{\mathrm{RpK}}_{\mathrm{bK}}$ for six
        bins of the Kp invariant mass with equal widths (lower figure). The coefficient $a_0$
        is close to 1, it is given in fig.~\ref{invKp-slices-fits}. The upper
        small figure shows the corresponding reduced $\chi^{2}$ of the fit. The dashed lines
        are generated with a second degree polynomial fit and have the only
        purpose to guide the eye.}
\end{figure}

\begin{figure}[htbp]
	\centering
	\includegraphics[width=.5\textwidth]{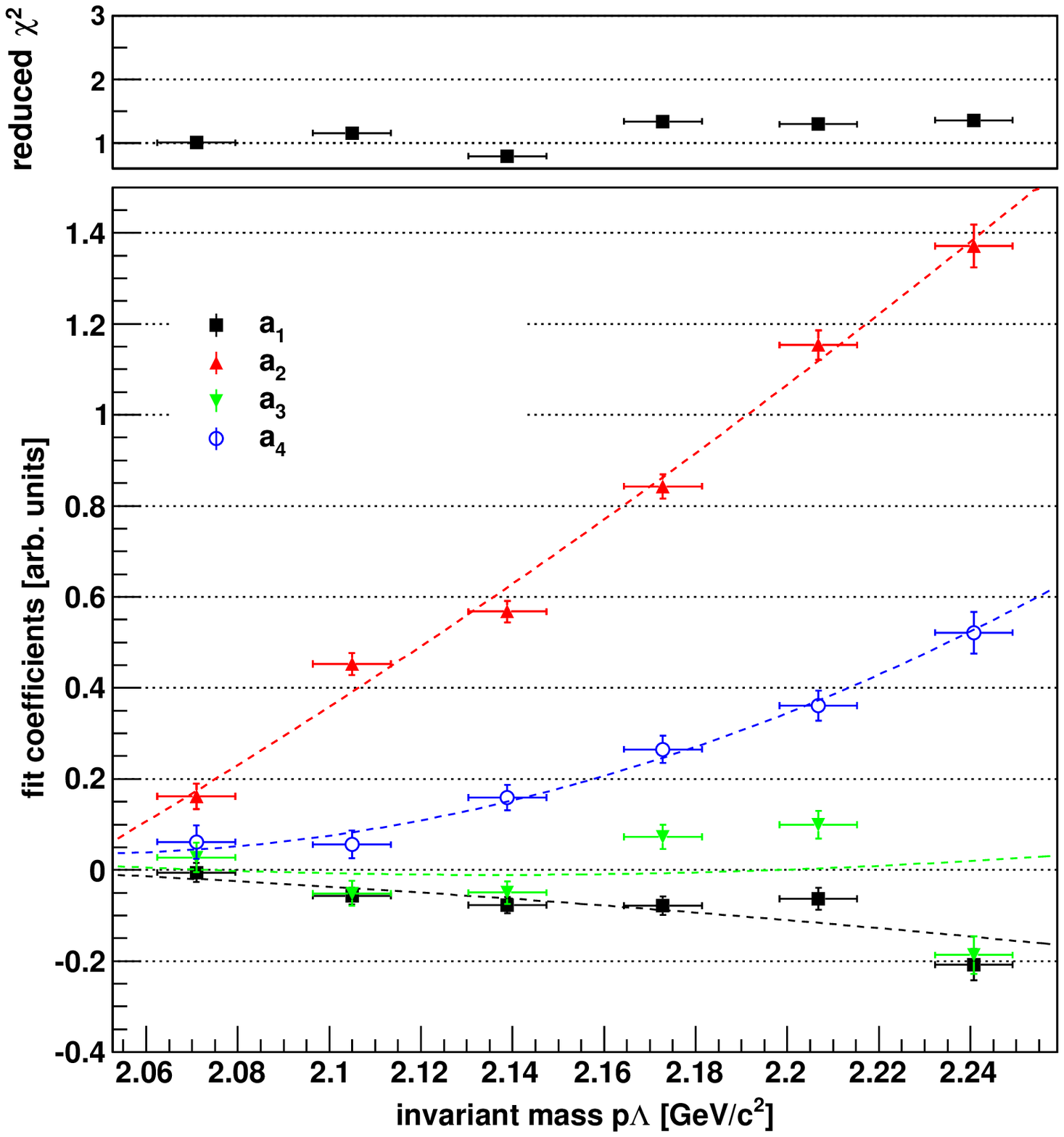}
	\caption{\label{invpL-coefficients}
         Fit results of the coefficients of equation \ref{legendrePolynomial} obtained
        from the GJ angular distributions $\cos\vartheta^{\mathrm{R\Lambda p}}_{\mathrm{bp}}$ for six
        bins of the p$\mathrm{\Lambda}$ invariant mass with equal widths (lower figure). The coefficient $a_0$
        is close to 1, it is given in fig.~\ref{invpL-slices-fits}. The upper
        small figure shows the corresponding reduced $\chi^{2}$ of the fit. The dashed lines
        are generated with a second degree polynomial fit  and have the only
        purpose to guide the eye.}
\end{figure}

As the K$\mathrm{\Lambda}$ invariant mass is proportional to the K and
$\mathrm{\Lambda}$ cm momentum the rising momenta of the
K$\mathrm{\Lambda}$ invariant mass slices in addition influences the
angular distributions, as it is shown in section
\ref{detailedAngularDistribution} for the cm angular
distributions. Therefore, an influence of N$^*$ resonances on the
variation of the fit coefficients $a_{0}$ .. $a_{4}$ with the
invariant mass can only be visible in the comparison with the
corresponding fit coefficients calculated in the same manner for the
invariant masses of pK and p$\mathrm{\Lambda}$, where no resonances
are expected. The GJ angular distributions for these invariant masses
are shown in the appendix (figs. \ref{invKp-slices-fits},
\ref{invpL-slices-fits}) and the variation of the coefficients in
figs. \ref{invKp-coefficients}, \ref{invpL-coefficients}.


No specific behavior of the angular momentum composition of the
K$\mathrm{\Lambda}$ invariant mass is detected, which could be related
to nucleon resonances. Therefore, this distribution seems to be the
wrong tool to detect the influence of nucleon resonances. The
invariant mass spectra, which contain the $\mathrm{\Lambda}$, which is
the heaviest particle of the reaction in study, show both a nearly
linear increase of the $a_2$ coefficient with the invariant mass and a
nearly quadratic increase of the $a_4$ coefficient.  For all three
invariant masses the odd coefficients are close to zero, thus
indicating that the angular distributions are almost symmetric.
%======================================================================================================================================
\section{Summary}
More than 200\,000 events of the reaction pp $\rightarrow$
pK$\mathrm{\Lambda}$ with a beam momentum of
$2.95\,\mathrm{GeV}/\mathrm{c}$ are analyzed. The angular
distributions of the final state particles in the cm system exhibit
different characteristics.  They are analyzed in dependence on cm
momentum by a fit of Legendre polynomials.  While the Legendre
coefficient $a_2$ of the $\mathrm{\Lambda}$ distribution rises with
the third power of the cm momentum, the coefficient $a_2$ of the
proton distribution rises linear with the cm momentum and is
approximately constant for values larger than
$0.35\,\mathrm{GeV}/\mathrm{c}$. The angular distribution of the kaon
is nearly constant for all cm momenta.

The threshold effect in the p$\mathrm{\Lambda}$ invariant mass
spectrum at the N$\mathrm{\Sigma}$ threshold can be described by a
Flatt\'{e} distribution by assuming a virtual state of
p$\mathrm{\Lambda}$ with a mass above the threshold. It is shown by
cuts on the pK helicity angle, that the height and the structure of
the cusp is nearly constant along the K$\mathrm{\Lambda}$ invariant
mass.  Therefore, the intensity variations are most likely due to
N$^*$ resonances decaying into K$\mathrm{\Lambda}$.  The height of the
cusp is isotropic in dependence on the kaon cm angle and in dependence
on the Gottfried-Jackson angle in the p$\mathrm{\Lambda}$ rest frame.

In the K$\mathrm{\Lambda}$ invariant mass no deviations at the
K$\mathrm{\Sigma}$ threshold are found. The upper-limit of a possible
threshold effect is estimated to be $\SI{0.3}{\micro\barn}$.  The
examination of the dependence of the Gottfried-Jackson angle in the
K$\mathrm{\Lambda}$ frame on the K$\mathrm{\Lambda}$ invariant mass
compared to the ones of the p$\mathrm{\Lambda}$ and Kp frames exhibits
no peculiarities. This indicates that this examination may not be
sensitive to effects of N$^*$ resonances.

\begin{acknowledgement}
The authors want to thank the COSY crew for the excellent beam preparation,
J. Uehlemann and N. Paul for the operation of the demanding LH$_2$ target. 
Discussions with J. Haidenbauer and C. Wilkin are gratefully acknowledged. 
This work was supported by grants from Forschungszentrum J\"{u}lich (COSY-FFE),
by the European Union Seventh Framework program (FP7/2007-2013) under
grant agreement 283286, and by the Foundation for Polish Science through the MPD programme.

\end{acknowledgement}
%======================================================================================================================================
\clearpage  % nur fuer referee mode
\section*{Appendix}
In the appendix the individual fits of the observables, which are separated into bins of a second observable, are shown.
Inside each figure the fit results and the range of the second observable are given.
\begin{figure*}[p]
	\centering
	\includegraphics[width=.95\textwidth]{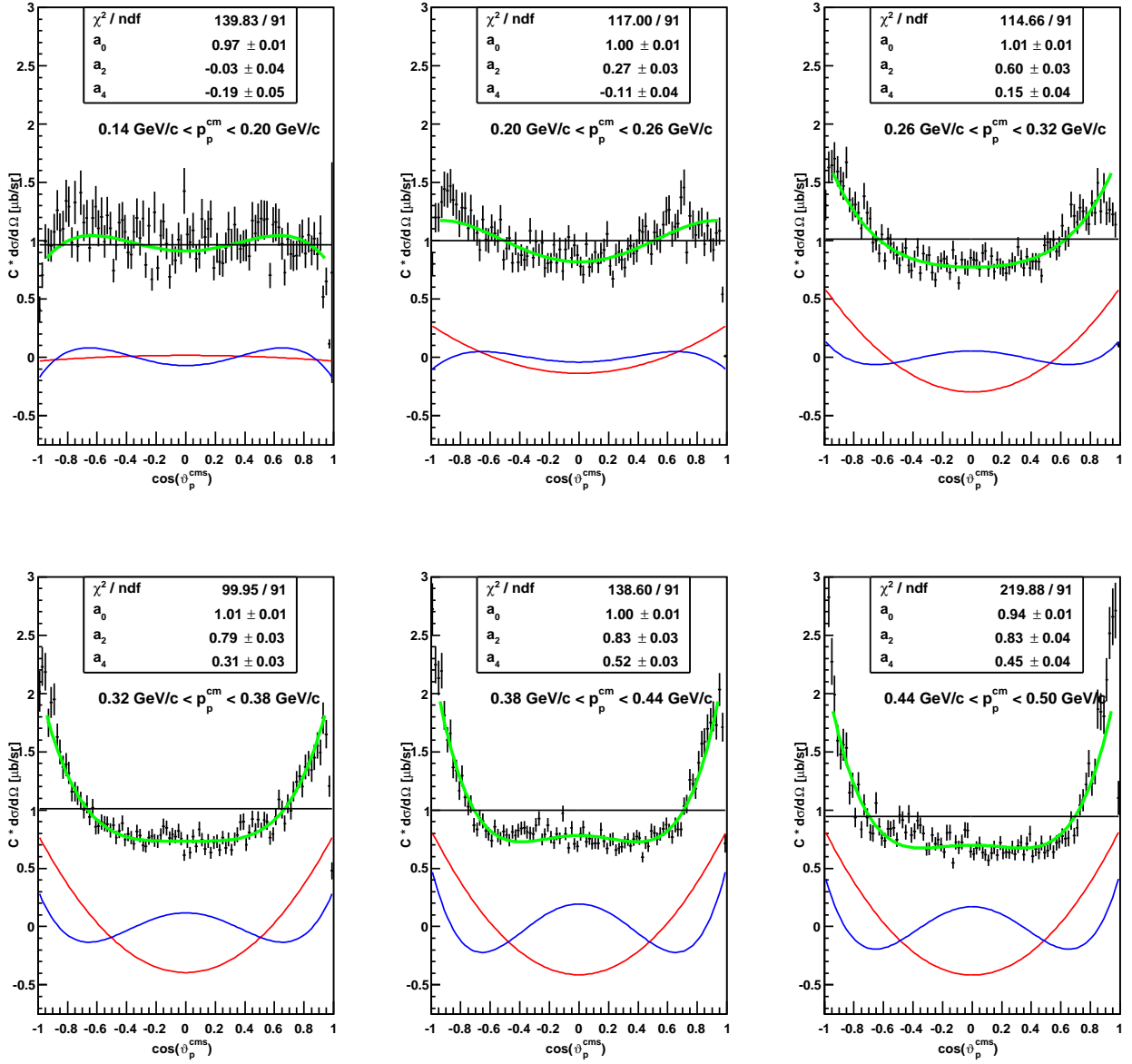}
	\caption{\label{ProtonAngularDistributions} 
          Proton angular distributions for different ranges of the proton 
        center of mass momentum. The green curve shows the fit result. The contributions
        of the Legendre polynomials  $P_{0}$, $P_{2}$, and $P_{4}$, weighted with the corresponding
        coefficient are plotted in black, red, and blue.
        }
\end{figure*}\clearpage
\clearpage
\clearpage

\begin{figure*}[p]
	\centering
	\includegraphics[width=.95\textwidth]{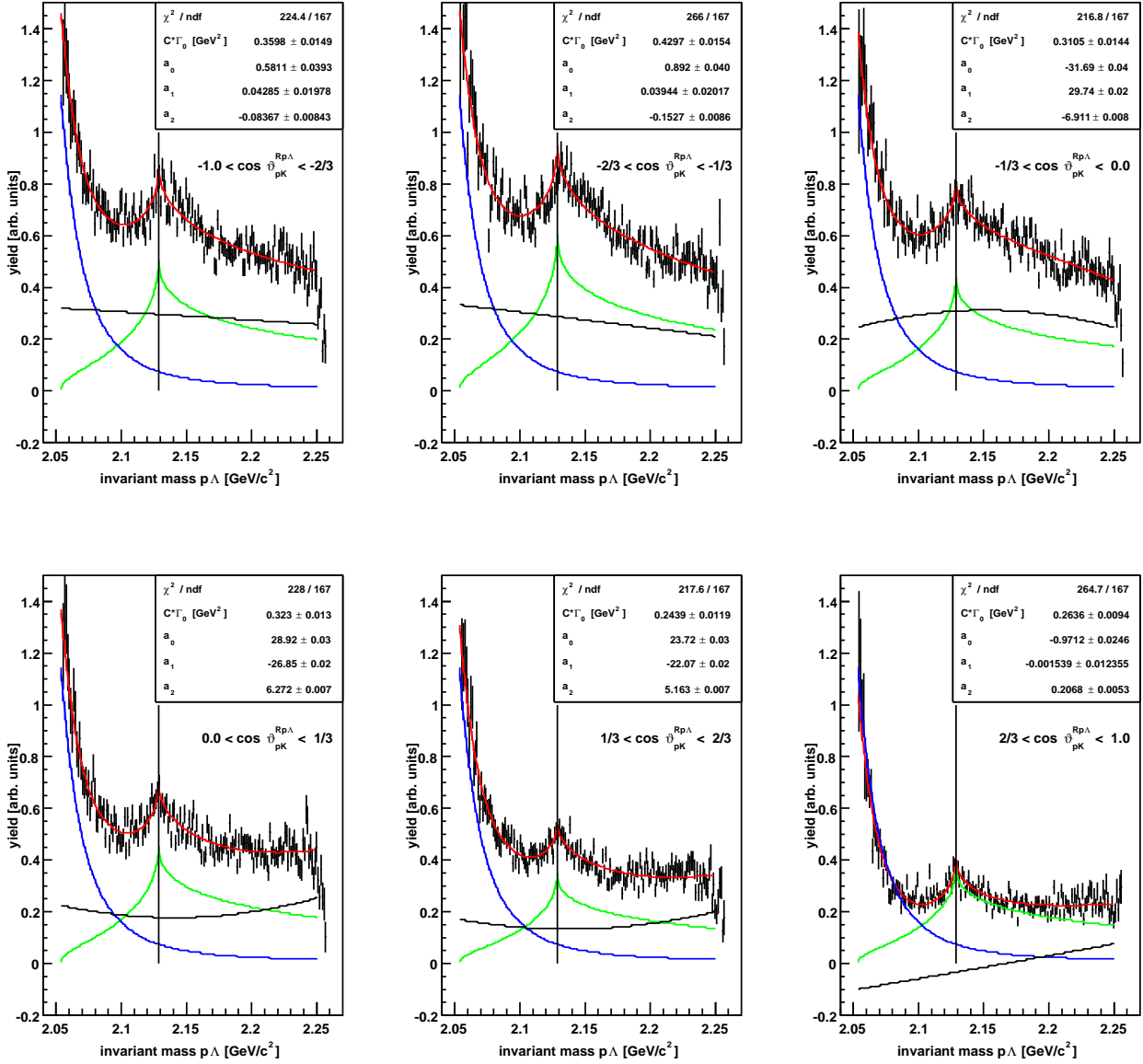}
	\caption{\label{helicity-slices-fits} The p$\mathrm{\Lambda}$ invariant
          mass distributions are shown for different slices of the
          Dalitz plot. The slices are generated with cuts in the
          helicity angle cos$\vartheta^{\mathrm{Rp\Lambda}}_{\mathrm{pK}}$ as
          given in the individual plots. The spectra are normalized to
          1/6 of the total data set which is shown in
          fig.~\ref{mpl-gesamt-Fit}.  The data are shown as points
          with error bars, the red line is the total fit.
          The green
          line is the contribution of the Flatt\'{e} distribution, where
          only the height ($C\!\Gamma_0$) is varied.
          The blue line is the  FSI part with all parameters
          fixed. The black line is the part of the reflections with freely
          varying parameters of the second order polynomial.  }
\end{figure*}\clearpage

\clearpage
\begin{figure*}[p]
	\centering
	\includegraphics[width=.95\textwidth]{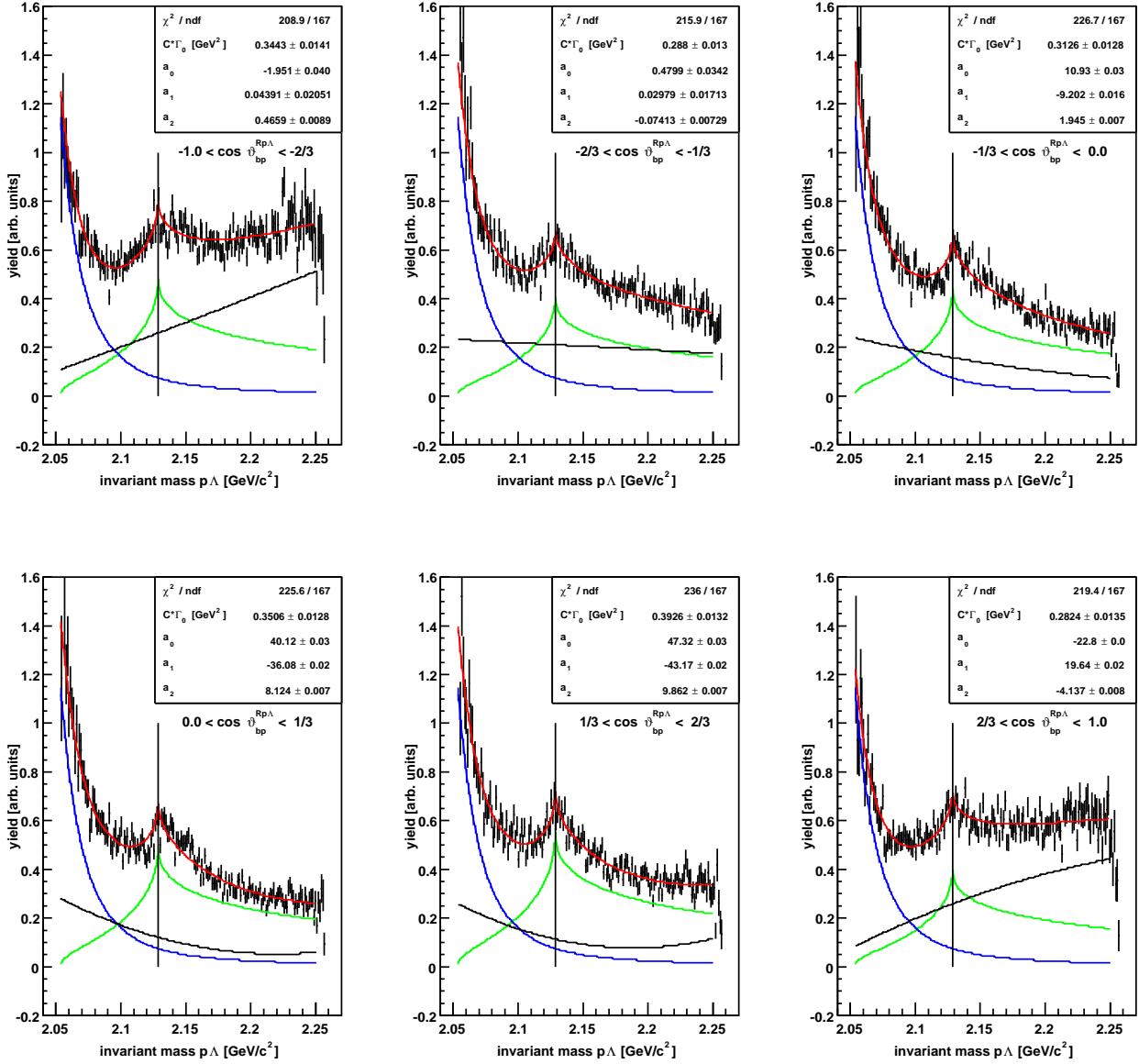}
	\caption{\label{gfj-slices-fits} The p$\mathrm{\Lambda}$ invariant
          mass distributions are shown for different slices
          of the proton - beam  angle in the GJ p$\mathrm{\Lambda}$ system as given in the individual plots.
          The spectra are normalized to
          1/6 of the total data set which is shown in
          fig.~\ref{mpl-gesamt-Fit}.  The data are shown as points
          with error bars, the red line is the total fit.   The green
          line is the contribution of the Flatt\'{e} distribution, where
          only the height ($C\!\Gamma_0$) is varied.
          The blue line is the  FSI part with all parameters
          fixed. The black line is the part of the reflections with freely
          varying parameters of the second order polynomial. }
\end{figure*}\clearpage
\begin{figure*}[p]
	\centering
	\includegraphics[width=.95\textwidth]{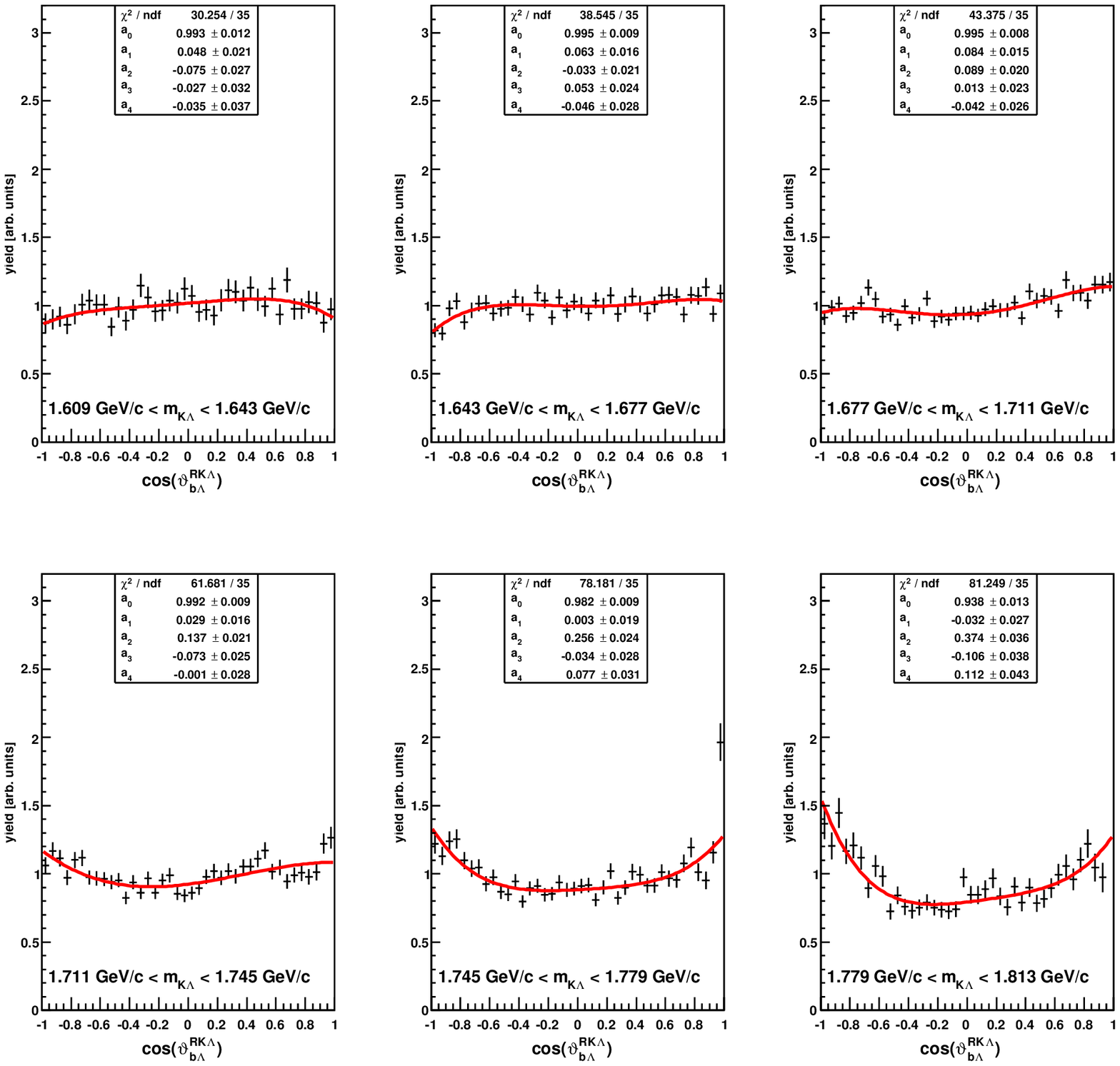}
	\caption{\label{invKL-slices-fits}  
        The GJ angular distribution cos $\vartheta^{\mathrm{RK\Lambda}}_{\mathrm{b\Lambda}}$ is shown for
        six slices of the K$\mathrm{\Lambda}$ invariant mass, the invariant mass ranges of the slices are given
        in the figures. The data, which are corrected to the acceptance and reconstruction efficiency, are shown by black error bars
        and the fit result according to equation \ref {legendrePolynomial} is shown by the red line. }
\end{figure*}\clearpage
\begin{figure*}[p]
	\centering
	\includegraphics[width=.95\textwidth]{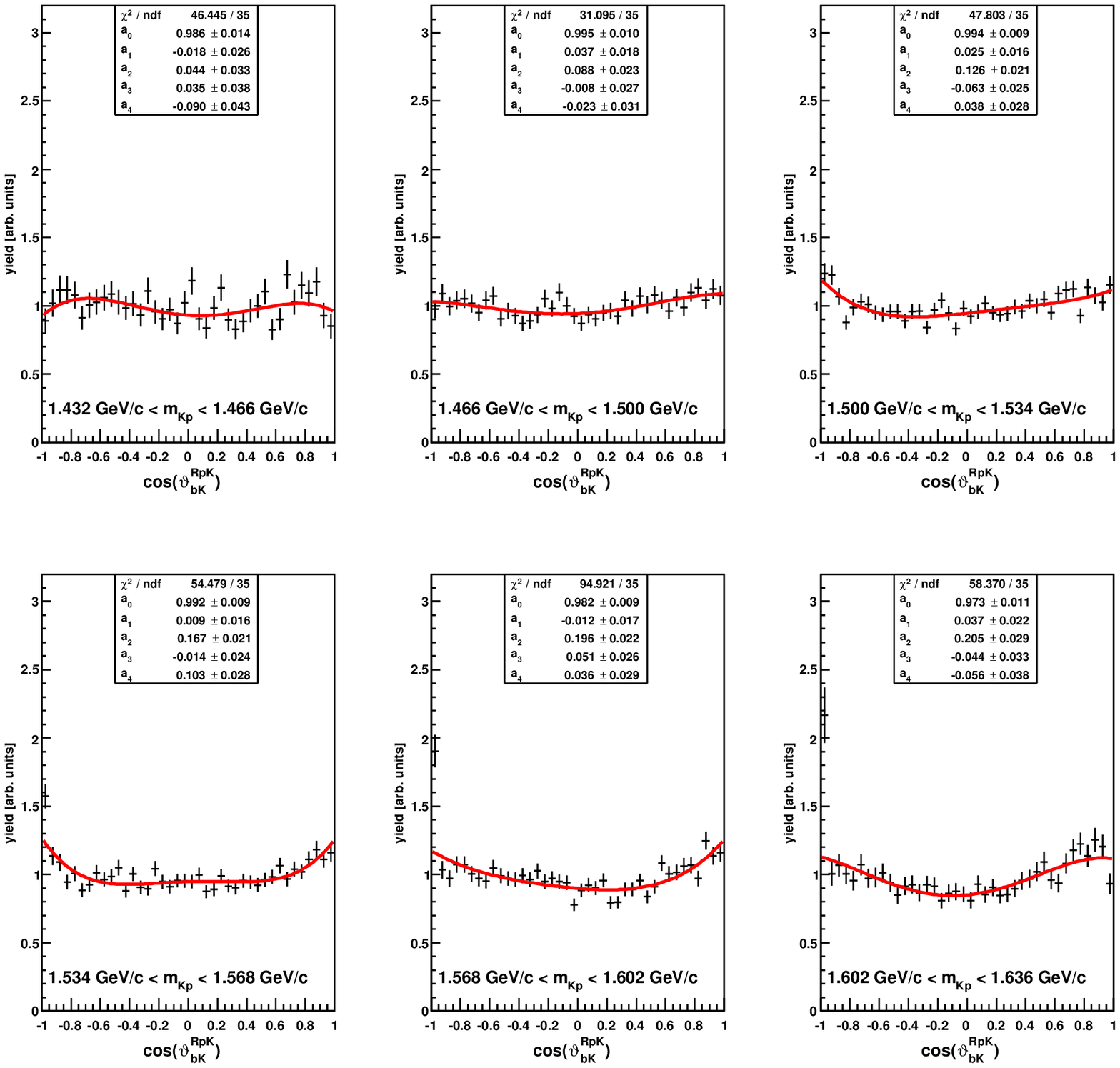}
	\caption{\label{invKp-slices-fits} The GJ angular distribution cos $\vartheta^{\mathrm{RpK}}_{\mathrm{bK}}$ is shown for
        six slices of the Kp invariant mass, the invariant mass ranges of the slices are given
        in the figures. The data, which are corrected to the acceptance and reconstruction efficiency, are shown by black error bars
        and the fit result  according to equation \ref {legendrePolynomial} is shown by the red line.  }
\end{figure*}\clearpage
\begin{figure*}[p]
	\centering
	\includegraphics[width=.95\textwidth]{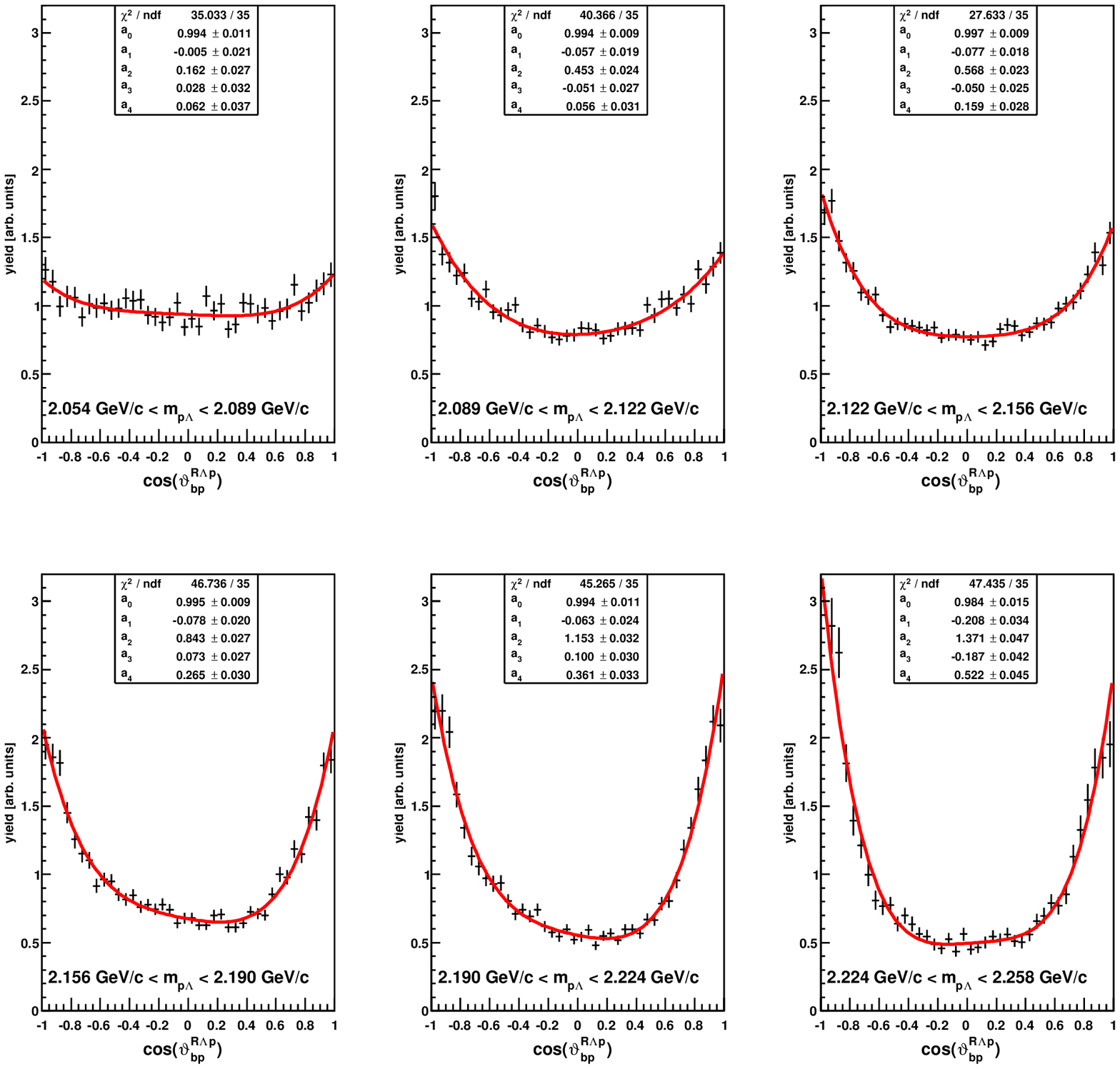}
	\caption{\label{invpL-slices-fits} The GJ angular distribution cos $\vartheta^{\mathrm{R\Lambda p}}_{\mathrm{bp}}$ is shown for
        six slices of the p$\mathrm{\Lambda}$ invariant mass, the invariant mass ranges of the slices are given
        in the figures. The data, which are corrected to the acceptance and reconstruction efficiency, are shown by black error bars
        and the fit result according to equation \ref {legendrePolynomial} is shown by the red line.  }
\end{figure*}\clearpage
%-----------------------------------------------------------------------------------------------------------------------------

\bibliographystyle{epj} 
\bibliography{references_pKL295}

\end{document}